\documentclass[fleqn,usenatbib]{mnras}

\usepackage{newtxtext,newtxmath}

\usepackage[T1]{fontenc}
\usepackage{ae,aecompl}

\newcommand{\corot}{CoRoT}

\newcommand{\gaia}{{\it Gaia}}

\newcommand{\spitzer}{{\it Spitzer}}
\newcommand{\panstarrs}{Pan-STARRS}
\newcommand{\gpe}{{GP-EBOP}}

\newcommand{\nex}{0.7 ex}

\newcommand{\kms}{km\,s$^{-1}$}

\newcommand{\msun}{\mbox{M$_{\odot}$}}
\newcommand{\rsun}{\mbox{R$_{\odot}$}}
\newcommand{\teff}{$T_{\rm eff}$}
\newcommand{\logg}{$\log g$}
\newcommand{\logL}{$\log L$}

\newcommand{\ebv}{E(B-V)}

\newcommand{\name}{Mon-735}
\newcommand{\ebdisk}{CoRoT\,223992193}
\newcommand{\cluster}{NGC\,2264}
\newcommand{\usco}{Upper Scorpius}

\newcommand{\colblue}[1]{\textcolor{blue}{#1}}

\newcommand{\bp}{G$_{\rm BP}$}
\newcommand{\rp}{G$_{\rm RP}$}

\newcommand{\ks}{K$_{\rm s}$}
\newcommand{\Sone}{IRAC-1}
\newcommand{\Stwo}{IRAC-2}
\newcommand{\um}{$\mu$m}

\newcommand{\mrp}{mass--radius}
\newcommand{\tlp}{\teff--\logL}

\newcommand{\btsettl}{BT-Settl}
\newcommand{\phoenix}{PHOENIX}

\newcommand{\psize}{0.4}

\newcommand{\Mpri}{0.2918\,$\pm$\,0.0099}
\newcommand{\Msec}{0.2661\,$\pm$\,0.0095}
\newcommand{\Rpri}{0.762\,$\pm$\,0.022}
\newcommand{\Rsec}{0.748\,$\pm$\,0.023}
\newcommand{\Tpri}{3260\,$\pm$\,73}
\newcommand{\Tsec}{3213\,$\pm$\,73}
\newcommand{\period}{1.9751388\,$\pm$\,0.0000050}
\newcommand{\dist}{718\,$\pm$\,27}
\newcommand{\distnoplxprior}{715\,$\pm$\,29}
\newcommand{\teffBTPHdiff}{160}
\newcommand{\distdiff}{90}

\newcommand{\Mpriebd}{0.665\,$\pm$\,0.015}
\newcommand{\Msecebd}{0.486\,$\pm$\,0.012}
\newcommand{\Rpriebd}{$1.291\,^{+0.031}_{-0.036}$}
\newcommand{\Rsecebd}{$1.118\,^{+0.042}_{-0.039}$}
\newcommand{\Tpriebd}{$3680\,^{+58}_{-52}$}
\newcommand{\Tsecebd}{$3645\,^{+58}_{-53}$}
\newcommand{\distebd}{$735\,^{+51}_{-37}$}
\newcommand{\distebdnoplxprior}{$733\,^{+57}_{-41}$}
\newcommand{\ebdteffBTPHdiff}{125}
\newcommand{\ebddistdiff}{80}

\usepackage{graphicx}	
\usepackage{amsmath}	
\usepackage{amssymb}	
\usepackage{multirow}
\usepackage{xcolor}

\title[A new low-mass PMS EB in NGC\,2264]{\name: A new low-mass pre-main sequence eclipsing binary in NGC\,2264}

\author[E. Gillen et al.]{
\parbox{\textwidth}{
Edward Gillen,$^{1,\dagger}$\thanks{E-mail: ecg41@cam.ac.uk (EG)}
Lynne A. Hillenbrand$^{2}$,
John Stauffer$^{3}$,
Suzanne Aigrain$^{4}$, \\
Luisa Rebull$^{5}$
and Ann Marie Cody$^{6}$ 
\vspace{2mm}} \\
$^{1}$Astrophysics Group, Cavendish Laboratory, J.J. Thomson Avenue, Cambridge CB3 0HE, UK \\ 
$^{2}$Department of Astronomy, California Institute of Technology, Pasadena, CA 91125, USA \\
$^{3}$Spitzer Science Center, California Institute of Technology, 1200 E California Blvd., Pasadena, CA 91125, USA \\
$^{4}$Sub-department of Astrophysics, Department of Physics, University of Oxford, Keble Road, Oxford, OX1 3RH, UK \\
$^{5}$Infrared Science Archive (IRSA), IPAC, 1200 E.\ California Blvd., California Institute of Technology, Pasadena, CA 91125, USA \\
$^{6}$NASA Ames Research Center, Moffet Field, CA 94035, USA \\
$^{\dagger}$Winton Fellow
}
\date{Accepted 2020 April 3. Received 2020 March 31; in original form 2019 August 14}

\pubyear{2020}

\begin{document}
\label{firstpage}
\pagerange{\pageref{firstpage}--\pageref{lastpage}}
\maketitle

\begin{abstract}

We present \name, a detached double-lined eclipsing binary (EB) member of the $\sim$3 Myr old \cluster\ star forming region, detected by \spitzer. We simultaneously model the \spitzer\ light curves, follow-up Keck/HIRES radial velocities, and the system's spectral energy distribution to determine self-consistent masses, radii and effective temperatures for both stars. We find that \name\ comprises two pre-main sequence M dwarfs with component masses of $M$ = \Mpri\ and \Msec\ M$_{\odot}$, radii of $R$ = \Rpri\ and \Rsec\ R$_{\odot}$, and effective temperatures of \teff\ = \Tpri\ and \Tsec\ K. The two stars travel on circular orbits around their common centre of mass in $P$ = \period\ days. We compare our results for \name, along with another EB in \cluster\ (\ebdisk), to the predictions of five stellar evolution models. These suggest that the lower mass EB system \name\ is older than \ebdisk\ in the \mrp\ diagram (MRD) and, to a lesser extent, in the Hertzsprung--Russell diagram (HRD). The MRD ages of \name\ and \ebdisk\ are $\sim$7--9 and 4--6 Myr, respectively, with the two components in each EB system possessing consistent ages.

\end{abstract}




\section{Introduction}
\label{sec:intro}

Accurately predicting the fundamental properties of stars (e.g. masses, radii, effective temperatures, luminosities and ages) is a central requirement for observational astrophysics. Modern stellar evolution theory, for example, underpins our ability to accurately estimate: 
the initial mass function \citep[e.g.][]{Bastian10};
determine cluster and association ages \citep[e.g.][]{Soderblom14,Bell15}
and hence protoplanetary disk lifetimes and giant planet formation timescales \citep[e.g.][]{Alexander08,Williams11,Ribas15};
infer exoplanet parameters \citep[e.g.][]{Gaidos13,Jones16,Raynard18,Rickman19};
and constrain the age-activity-rotation relation \citep[e.g.][]{Barnes05,Meibom09,Meibom15}.
Uncertainties in model-derived stellar properties translate into systematic uncertainties in these areas. Significant effort is therefore directed into observationally testing stellar evolution models, especially at low masses (M\,<\,0.8\,\msun) and young ages (t\,<\,1 Gyr), as current constraints remain scarce.

Detached, double-lined eclipsing binary (EB) stars offer a useful test of stellar models, as the masses and radii of both stars can be directly measured, with minimal theoretical assumptions, from the light and radial velocity curves of the system. With high quality data and modern software tools, fundamental stellar parameters can be measured to precisions of 1--3\,\%; these provide one of the strongest observational tests on stellar evolution theory available \citep{Andersen91,Torres10,Stassun14}. Furthermore, with light curves in two or more (ideally well-separated) photometric bands, effective temperature (\teff) ratios can be determined from relative eclipse depths, albeit it with some dependence on either stellar atmosphere models or empirical relations (and to obtain absolute \teff's would require one component to be determined independently). Nonetheless, by simultaneously analysing photometric eclipses (ideally in multiple bands), radial velocities, and either spectral energy distributions or flux-calibrated spectra (or by performing spectral disentangling), it is possible to self-consistently solve for the stellar masses, radii, effective temperatures, and luminosities. Finally, parallax measurements (e.g. from \gaia; \citealt{Brown18}) give independent distance estimates that can be incorporated into the fit.

Open clusters are stellar populations that share the same age and metallicity, and thereby represent powerful tests of stellar evolution theory due to their ensemble nature. Young open clusters offer a window onto the early evolution of stellar and planetary systems and, as their member stars can be dated to a precision unattainable for young stars in the field, constrain timescales for the processes driving their evolution. Consequently, many photometric monitoring programs have targeted young open clusters to:
study accretion and the star--disk interaction \citep[e.g.][]{Bouvier07,Alencar10,Cody14,Stauffer14,Stauffer15};
understand the evolution of stellar rotation \citep[e.g.][]{Irwin06,Meibom09,Hartman10,Affer13,Rebull16,Rebull16a,Stauffer16,Douglas17,Gillen20} 
and pulsations \citep[e.g.][]{Zwintz06,Zwintz11}; 
and to search for young transiting planets  \citep[e.g.][]{David16b,Mann17,Pepper17,Livingston18}, 
brown dwarfs \citep[e.g.][]{Gillen17a,David19}, 
and eclipsing binaries \citep[e.g.][]{Stassun04,Irwin07,Gillen14,Kraus15,Kraus17,David16a}.

EBs in open clusters are particularly valuable because their metallicity can be inferred through their membership, and their derived age acts as an independent age estimate for the cluster. Multiple EBs in the same cluster, which share the same cluster age and metallicity, are even more valuable as they can act as a joint test on stellar evolution theory \citep[e.g.][]{Gillen17a,David19}. The search for new EBs in young open clusters was one of the primary science goals behind the \corot\ space mission \citep{Baglin03} observing the \cluster\ star forming region for 23 days in 2008, as well as subsequent simultaneous \corot\ and \spitzer\ \citep{Werner04} observations of \cluster\ in 2011--2012 (39 and 29 days, respectively). These observations detected several tens of EBs, most of which are field-age systems, with \citet{Gillen14} presenting the first confirmed low-mass EB member of the cluster. Here, we focus on the second low-mass EB member of \cluster, \name, which was detected through the 2011--2012 \spitzer\ observations. The position of \name\ was not covered by the aforementioned \corot\ observations.

The \cluster\ star forming region is the dominant component of the Mon OB1 association in the Monoceros constellation, which is situated in the local Orion--Cygnus spiral arm at a distance of $\sim$700--800 pc \citep{Sung97,Dahm08,Sung10,Gillen14}. \cluster\ is hierarchically structured with significant sub-clusterings and a halo population \citep[e.g.][]{Teixeira06}. It comprises $\sim$1500 stars and has a commonly accepted median age of $\sim$3 Myr with an apparent dispersion of $\sim$3--5 Myr \citep{Walker56,Park00,Rebull02}. This age dispersion arises from the fact that early type members (A0 and earlier) are believed to have reached the main sequence, which sets a minimum cluster age, and yet star formation is ongoing \citep{Young06,Dahm08,Sung10}. \cluster\ has been historically well-studied due to its large and well-defined membership, and low-foreground extinction of \ebv = 0.06--0.15 mag \citep[e.g.][]{Perez87,Rebull02,Mayne08}. The cluster radial velocity (RV) is formally estimated as RV = 22\,$\pm$\,3.5 \kms, but displays a significantly non-Gaussian dispersion where stars between RV = 8--36 \kms\ have been considered as potential members \citep{Furesz06,Tobin15}.

This paper presents the discovery and characterisation of \name\ as a young low-mass pre-main sequence (PMS) EB member of \cluster.  The source has been previously identified as a T Tauri type member of \cluster\ based on ultraviolet excess and/or optical variability  \citep{Rebull02,Lamm04,Barentsen13,Venuti14} and is listed as an EB in the surrounding \corot\ SRa01 field \citep{Klagyivik13}.
In \S \ref{sec:obs} we introduce the photometric and spectroscopic observations. 
In \S \ref{sec:analysis} we model the light curves, radial velocities and spectral energy distribution of the system to determine masses, radii, temperatures and luminosities for both stars. 
We report our results in \S \ref{sec:results} along with a re-analysis of \ebdisk\ to obtain a self-consistent set of parameters for both systems. 
In \S \ref{sec:discussion} we compare our fundamental parameters for both stars with the predictions of five modern stellar evolution models before concluding in \S \ref{sec:conclusions}.

\section{Observations}
\label{sec:obs}

\begin{table}
  \centering
  \caption{Names and coordinates for \name}
  \label{tab:info}
  \begin{tabular}{llr}
    \hline
    \hline
    \noalign{\smallskip}
    Property & Value  &  Epoch  \\
    \noalign{\smallskip}
    \hline
    \noalign{\smallskip}
    Names  &  CSI Mon-000735  &  \\
           &  2MASS J06412734+0942003 & \\
           &  Gaia DR2 3326699526510395264 & \\
    \noalign{\smallskip}
    R.A.  &  \,\,\,\,\,06 41 27.342  &  J2000.0  \\
    Dec   &  +09 42 00.378  &  J2000.0  \\
    
    \noalign{\smallskip}
    \hline
 \end{tabular}
\end{table}

\begin{table*}
  \centering
  \caption{Broadband system photometry of \name.}
  \label{tab:photometry}
  \begin{tabular}{llll}
    \hline
    \hline
    \noalign{\smallskip}
    Band & Magnitude  &  Spectral flux density  &  Refs.  \\
      &  & (erg\,s$^{-1}$\,cm$^{-2}$\,\AA$^{-1}$) &  \\
    \noalign{\smallskip}
    \hline
    \noalign{\smallskip}

  SDSS $u$            &  $22.018\pm0.201$ AB  &  $(1.35\pm0.25) \times 10^{-17}$      &  (1,5,6) \\
  SDSS $g$            &  $19.850\pm0.016$ AB  &  $(5.732\pm0.084) \times 10^{-17}$    &  (1,5,6) \\
  SDSS $r$            &  $18.446\pm0.009$ AB  &  $(1.202\pm0.010) \times 10^{-16}$    &  (1,5,6) \\
  SDSS $i$            &  $16.892\pm0.005$ AB  &  $(3.415\pm0.016) \times 10^{-16}$    &  (1,5,6) \\
  SDSS $z$            &  $16.040\pm0.007$ AB  &  $(5.252\pm0.034) \times 10^{-16}$    &  (1,5,6) \\
  \noalign{\smallskip}
  \panstarrs1 $g$     &  $19.700\pm0.037$ AB  &  $(6.10\pm0.21) \times 10^{-17}$      &  (2,5,7) \\
  \panstarrs1 $r$     &  $18.449\pm0.010$ AB  &  $(1.181\pm0.011) \times 10^{-16}$    &  (2,5,7) \\
  \panstarrs1 $i$     &  $16.882\pm0.005$ AB  &  $(3.388\pm0.014) \times 10^{-16}$    &  (2,5,7) \\
  \panstarrs1 $z$     &  $16.177\pm0.002$ AB  &  $(4.895\pm0.010) \times 10^{-16}$    &  (2,5,7) \\
  \panstarrs1 $y$     &  $15.829\pm0.009$ AB  &  $(5.475\pm0.047) \times 10^{-16}$    &  (2,5,7) \\
  \noalign{\smallskip}
  \gaia\ $G$          &  $17.6210\pm0.0025$   &  $(2.2763\pm0.0052) \times 10^{-16}$  &  (3,5,8) \\
  \gaia\ \bp          &  $19.0371\pm0.0444$   &  $(1.008\pm0.041) \times 10^{-16}$    &  (3,5,8) \\
  \gaia\ \rp          &  $16.3417\pm0.0083$   &  $(3.825\pm0.029) \times 10^{-16}$    &  (3,5,8) \\
  \noalign{\smallskip}
  2MASS $J$           &  $14.491\pm0.029$     &  $(5.00\pm0.13) \times 10^{-16}$      &  (4,5,9) \\
  2MASS $H$           &  $13.910\pm0.025$     &  $(3.093\pm0.071) \times 10^{-16}$    &  (4,5,9) \\
  2MASS \ks           &  $13.520\pm0.039$     &  $(1.674\pm0.060) \times 10^{-16}$    &  (4,5,9) \\
  \noalign{\smallskip}
  \spitzer\ IRAC-1    &  $13.424\pm0.015$     &  $(2.873\pm0.040) \times 10^{-17}$    &  (This work,5,10) \\
  \spitzer\ IRAC-2    &  $13.342\pm0.015$     &  $(1.237\pm0.017) \times 10^{-17}$    &  (This work,5,10) \\

    \noalign{\smallskip}
    \hline
 \end{tabular}
 \begin{list}{}{}  
   \item[\textbf{Notes.}]{Magnitudes and spectral flux densities typically have formal measurement errors, which do not reflect the system's intrinsic variability. SDSS and \panstarrs\ magnitudes are in the AB system, while \gaia, 2MASS and \spitzer\ magnitudes are in Vega.}
     \item[\textbf{References.}] \emph{Photometry}: 1. \citet{Aguado19}; 2. \citet{Chambers16}; 3. \citet{Brown18}; 4. \citet{Skrutskie06}. \emph{Bandpasses}: 5. Filter Profile Service (FPS; \href{http://svo2.cab.inta-csic.es/theory/fps}{http://svo2.cab.inta-csic.es/theory/fps}); 6. SDSS DR7 (\href{https://classic.sdss.org/dr7/instruments/imager/\#filters}{https://classic.sdss.org/dr7/instruments/imager/\#filters}); 7. \citet{Tonry12}; 8. \citet{Evans18}; 9. \citet{Cohen03}; 10. NASA/IPAC Infrared Science Archive (\href{https://irsa.ipac.caltech.edu/data/SPITZER/docs/irac}{https://irsa.ipac.caltech.edu/data/SPITZER/docs/irac}).
  \end{list}
\end{table*}

\begin{table*}  
 \centering  
 \caption{Radial velocities and flux ratios derived from Keck/HIRES spectra.} 
 \label{tab:RVs}  
 \begin{tabular}{c c c c c r r}  
 \noalign{\smallskip} \hline  \hline \noalign{\smallskip} 
   UT date  &  HJD  &  Phase\,*  & S/N &  Light ratio  &  Primary RV  &  Secondary RV  \\
     &    &    &  (7550 \AA)  &  (6200--7200 \AA)  &  (km\,s$^{-1}$)  &  (km\,s$^{-1}$)  \\
\noalign{\smallskip} \hline \noalign{\smallskip}

2015-10-27  &  2457323.0845  &  0.81  &  18   &  $0.895\pm0.101$  &  $78.19\pm3.64$  &  $-48.04\pm1.53$  \\
2015-12-24  &  2457381.0684  &  0.17  &  21   &  $0.922\pm0.123$  &  $-40.48\pm3.11$  &  $80.98\pm2.84$  \\
2015-12-29  &  2457385.9827  &  0.65  &  25   &  $0.896\pm0.122$  &  $73.19\pm1.46$  &  $-44.27\pm2.99$  \\
2016-02-02  &  2457420.8102  &  0.28  &  16   &  $0.879\pm0.068$  &  $-46.39\pm1.98$  &  $91.45\pm2.53$  \\
2016-02-03  &  2457421.7907  &  0.78  &  11   &  $1.079\pm0.058$  &  $84.28\pm2.98$  &  $-52.03\pm3.28$  \\
2016-10-14  &  2457676.0175  &  0.48  &  13   &  \multicolumn{1}{c}{---}  &   \multicolumn{2}{c}{$18.58\pm1.04$}  \\
2016-12-22  &  2457744.9145  &  0.38  &  21   &  $0.968\pm0.131$  &  $-28.47\pm2.16$  &  $69.10\pm1.58$  \\
2017-01-13  &  2457766.8100  &  0.45  &  17   &  $0.865\pm0.132$  &  $3.64\pm1.93$  &  $35.32\pm1.92$  \\

 \noalign{\smallskip}
 \hline  
 \end{tabular}  
\begin{list}{}{}  
\item[*] Phase is defined relative to primary eclipse.   
\end{list}  
 \end{table*}

\begin{figure*}
   \centering
   \includegraphics[width=0.8\linewidth]{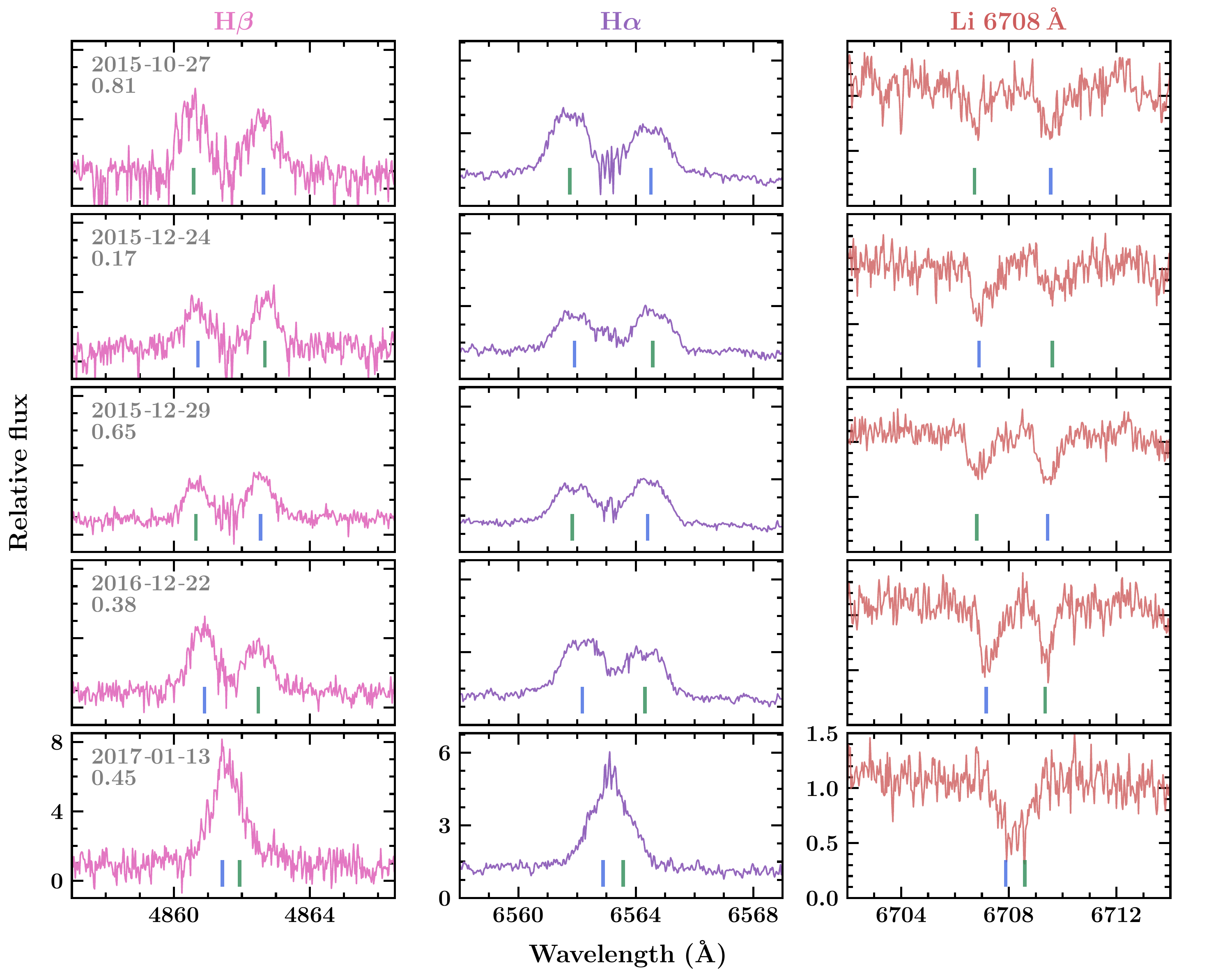}
    \caption{H$\beta$, H$\alpha$ and Li\,6708\,\AA\ profiles of \name\ (columns, left-to-right) for different epochs (rows, top-to-bottom). The epoch and orbital phase are given in the H$\beta$ plots. The vertical blue and green lines indicate the radial velocities of the primary and secondary stars, respectively, at each epoch. Both stars present H$\beta$ and H$\alpha$ emission profiles, and Li\,6708\,\AA\ absorption profiles, which suggest that \name\ is a chromospherically active young system.}
   \label{fig:spectra}
\end{figure*}

The \cluster\ star forming region was the subject of a photometric and spectroscopic monitoring campaign from 15 ground- and space-based telescope facilities in late 2011. This Coordinated Synoptic Investigation of NGC 2264 (CSI 2264) is introduced in detail in \citet{Cody14}. Here, we focus on the \spitzer/IRAC 3.6 and 4.5 \um\ observations (PI Stauffer), which discovered the \name\ EB system, and follow-up spectroscopic measurements with the Keck HIRES spectrograph. The names, coordinates and broadband magnitudes of \name\ are given in Tables \ref{tab:info} and \ref{tab:photometry}.

\subsection{Photometry}
\label{sec:phot}

\name\ was observed by \spitzer/IRAC at 3.6 and 4.5 \um\ for 28.7 days between 4 December 2011 and 1 January 2012 at a typical cadence of 1.7 hours. The 3.6 and 4.5 \um\ observations are near-simultaneous, with the 4.5 \um\ data taken $\sim$20 mins after the 3.6 \um\ observations. \spitzer\ performed four 20-hour continuous (stare mode) observations towards the beginning of the campaign, leading to corresponding data gaps in the photometry of \name, as it did not fall within the continuous observation region. The \spitzer\ data was converted from magnitudes into relative flux units for the analysis presented here. The reader is referred to \citet{Cody14} for further details of the \spitzer\ observations, data reduction and light curve production\footnote{See \href{https://irsa.ipac.caltech.edu/data/SPITZER/CSI2264/}{https://irsa.ipac.caltech.edu/data/SPITZER/CSI2264/} for the \spitzer\ data archive.}.

\begin{figure*}
   \centering
   \includegraphics[width=0.95\linewidth]{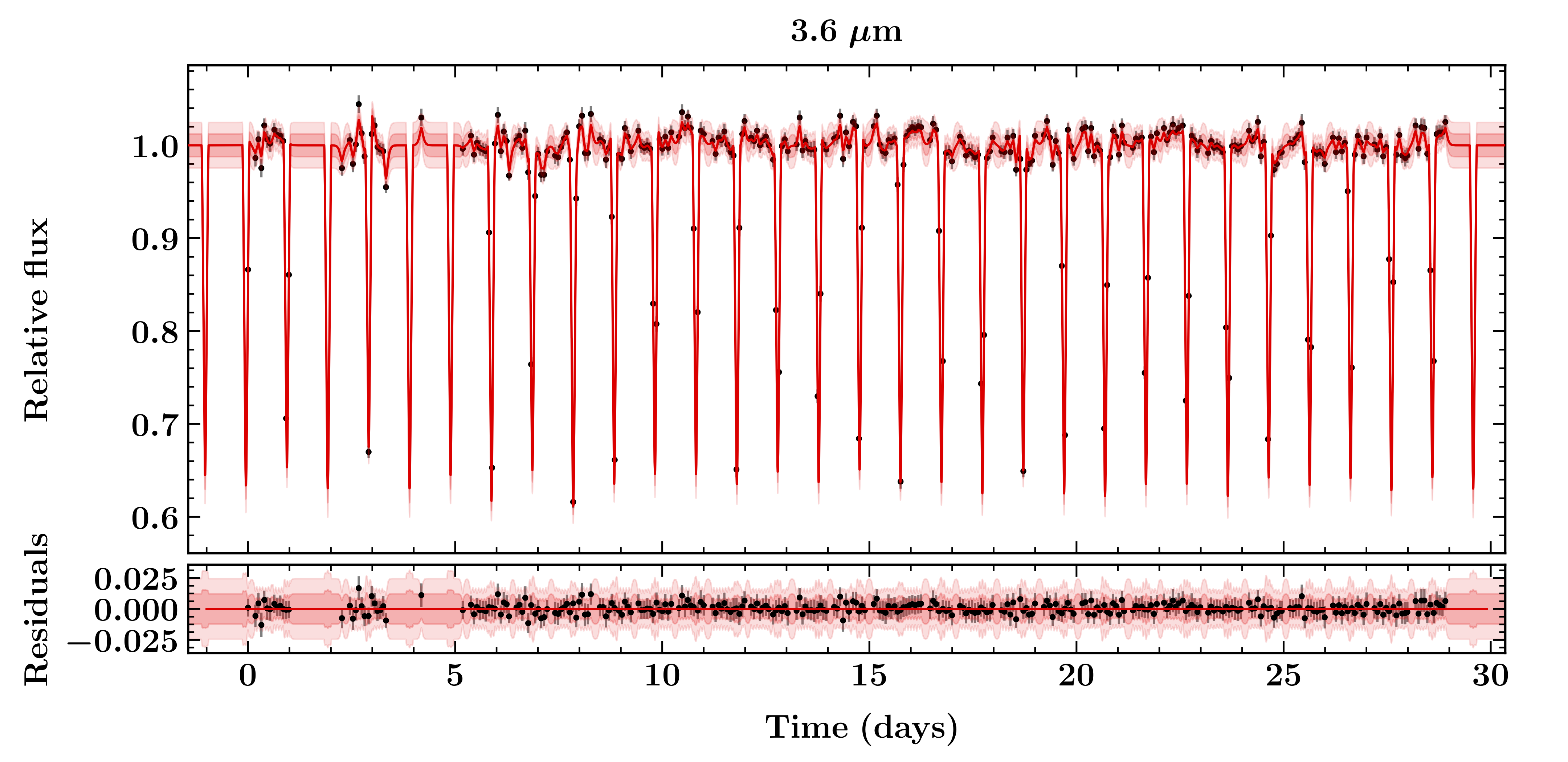}
   \includegraphics[width=0.95\linewidth]{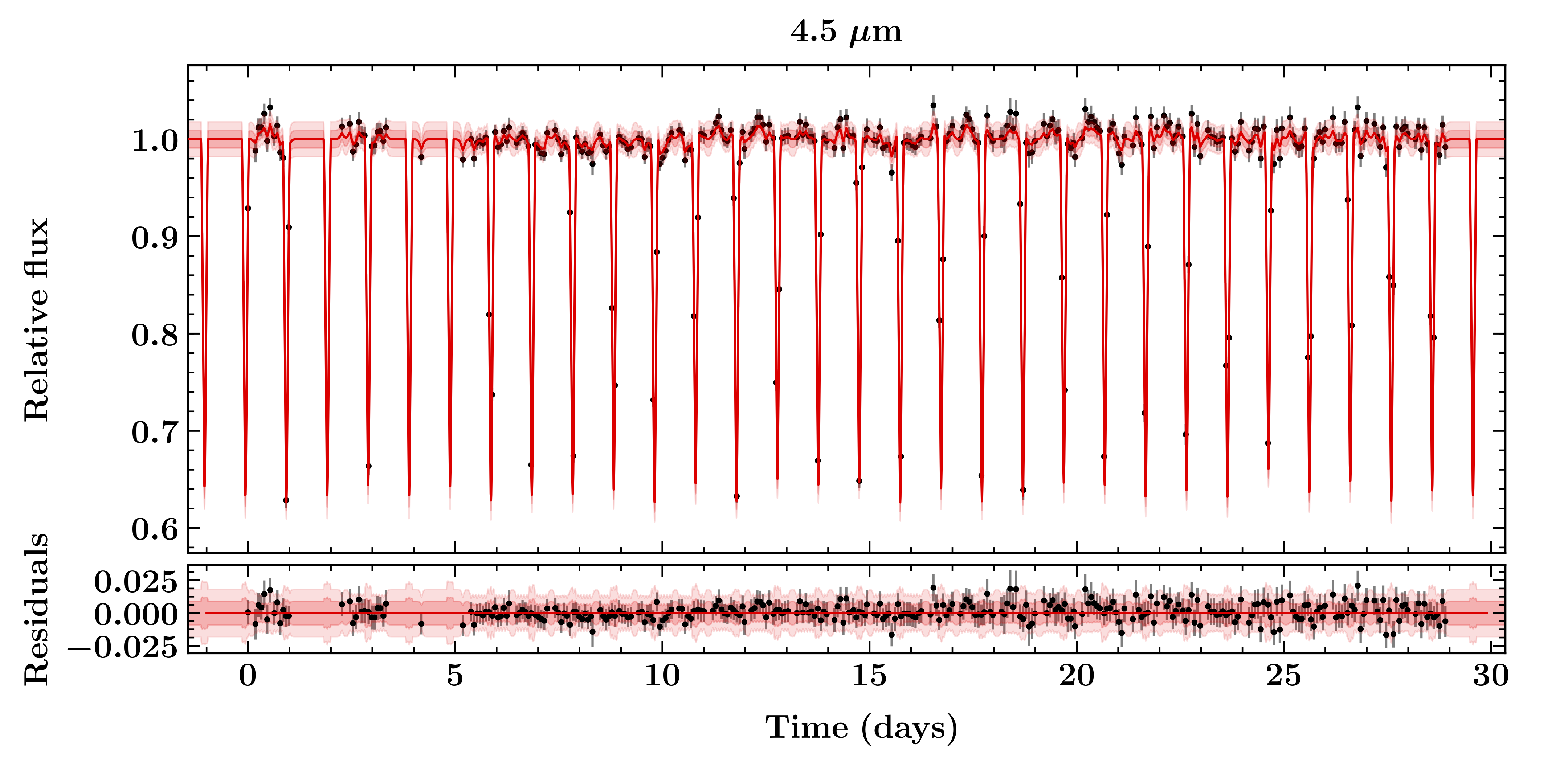}
    \caption{\spitzer\ IRAC 3.6 and 4.5 $\mu$m relative flux light curves of \name\ (black points, top and bottom, respectively) with the \gpe\ model in red and residuals below. In both light curves, the red line and pink shaded regions show the mean and 1 \& 2$\sigma$ confidence intervals of the predictive posterior distribution. The system displays near equal eclipses on both stars, which are $\sim$36--37\% deep, as well as relatively rough short-timescale out-of-eclipse variability.}
   \label{fig:LCs}
\end{figure*}

\subsection{Spectroscopy}
\label{sec:spec}

We obtained eight high-resolution optical spectra with the Keck HIRES spectrograph \citep{Vogt94} between 27 October 2015 and 13 January 2017. The HIRES spectra cover the wavelength range $\sim$4800--9200\,\AA\ with a resolving power of R\,$>$\,36,000,  and were reduced using the dedicated \textsc{makee} software written by Tom Barlow. 

Details of the individual observations are given in Table \ref{tab:RVs} along with information from the analysis presented \S \ref{sec:RVs}. The \name\ system is double-lined in the HIRES spectra at all epochs bar one (2016-10-14), where only a single peak was apparent, corresponding to both components of the system being located close to the systematic velocity. We determined a spectral type of M$3.1 \pm 0.7$ for this composite spectrum by measuring the strength of several TiO features relative to a set of spectral standards. The analytically determined spectral type was verified by visual examination of the spectrum details over a range of wavelengths.

The spectra display weak H$\alpha$ and H$\beta$ emission, as well as Li I\,6708\,\AA\ absorption, which together suggest \name\ is a chromospherically active young system \citep{stauffer1986,west2011,soderblom2014}. Examples of these three spectral features at different phases of the EB orbit are shown in Figure \ref{fig:spectra}.  
The H$\alpha$ equivalent width is about -2.5 \AA\ for each component of the EB, with line profiles that are flat-topped, as is often seen in chromospherically-active moderately-rotating mid-M type stars \citep[see examples in Appendix A of][]{ls2010}. The Ca II triplet lines are purely in absorption with no obvious central line reversal; thus neither component of the system seems to be highly active \citep{schofer2019}, or an accretor \citep{herbig1980}. The Li I\,6708\,\AA\  absorption strength is approximately 290 and 320 mA in the two components, which is typical for the pseudo-equivalent width of \cluster\ members at this spectral type, however somewhat low for truely measured equivalent widths that account for the significant TiO absorption in this spectral region \citep{bouvier2016}.

\section{Analysis with \gpe}
\label{sec:analysis}

\begin{figure*}
   \centering 
   \includegraphics[width=0.48\linewidth]{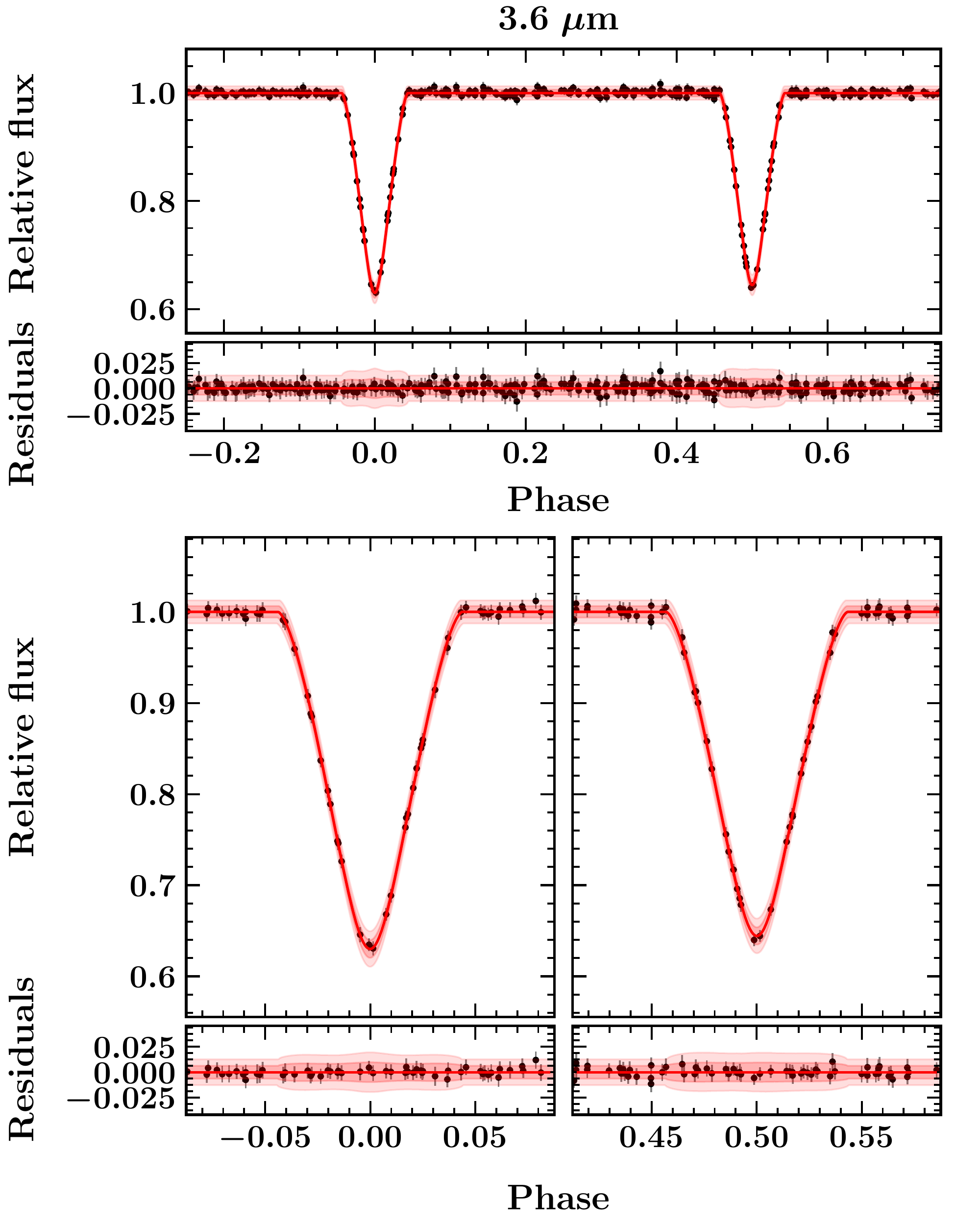}
   \hfill
   \includegraphics[width=0.48\linewidth]{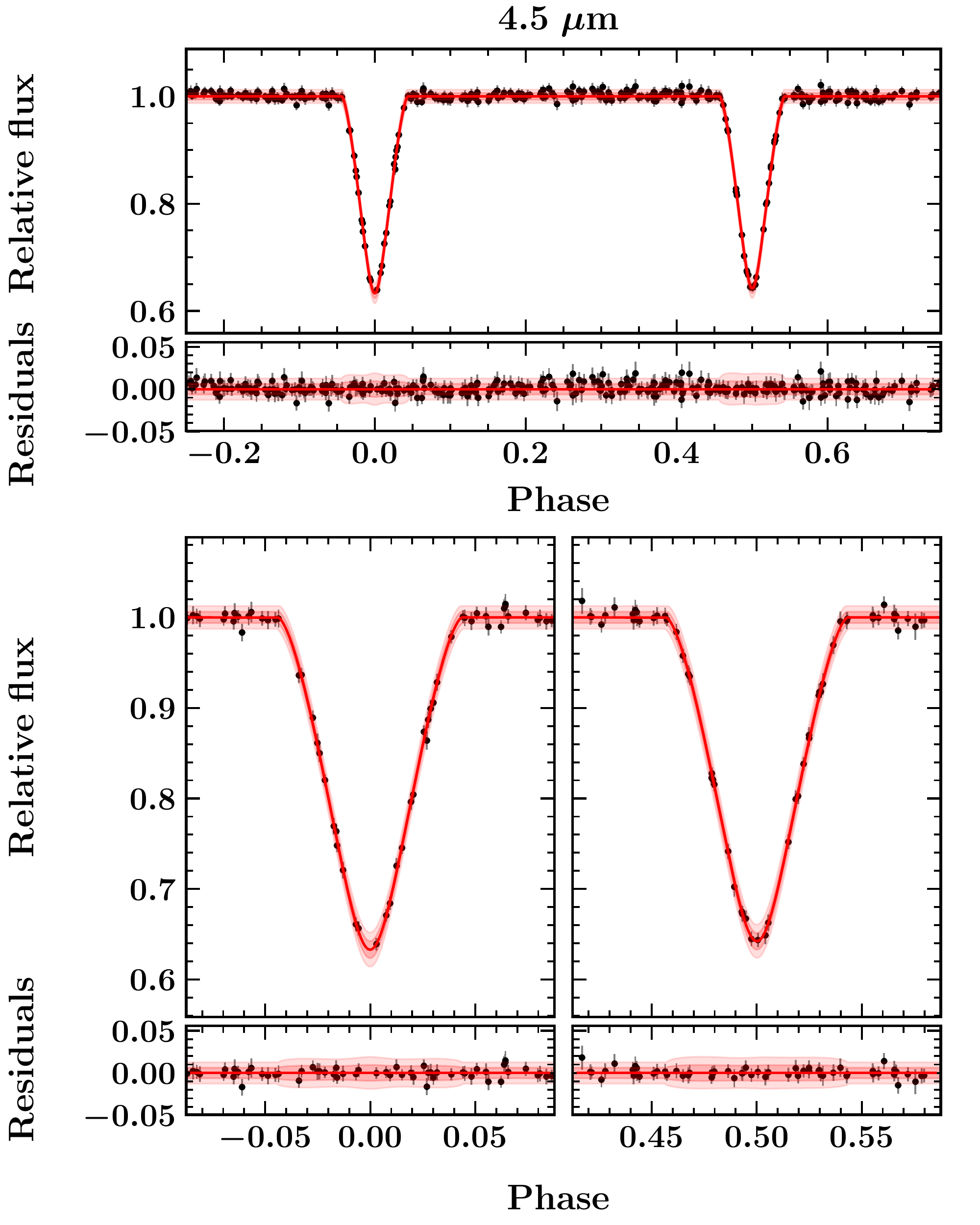}
    \caption{Phase-folded \spitzer\ IRAC 3.6 and 4.5 $\mu$m light curves (left and right, respectively). Top panels: full phase-folded light curves that have been detrended with respect to the Gaussian process model, with residuals immediately below. The red line represents the median posterior EB model, i.e. the median path of individual draws computed in phase space, and the pink shaded regions indicate the 1 \& 2$\sigma$ confidence intervals. The phase is defined relative to primary eclipse centre. Bottom panels: zooms on the primary and secondary eclipses (left and right, respectively, for each light curve), with residuals immediately below. The eclipses are grazing with near equal depths of $\sim$36--37\% on both stars.}
   \label{fig:eclipses}
\end{figure*}

\begin{figure}
   \centering
   \includegraphics[width=\linewidth]{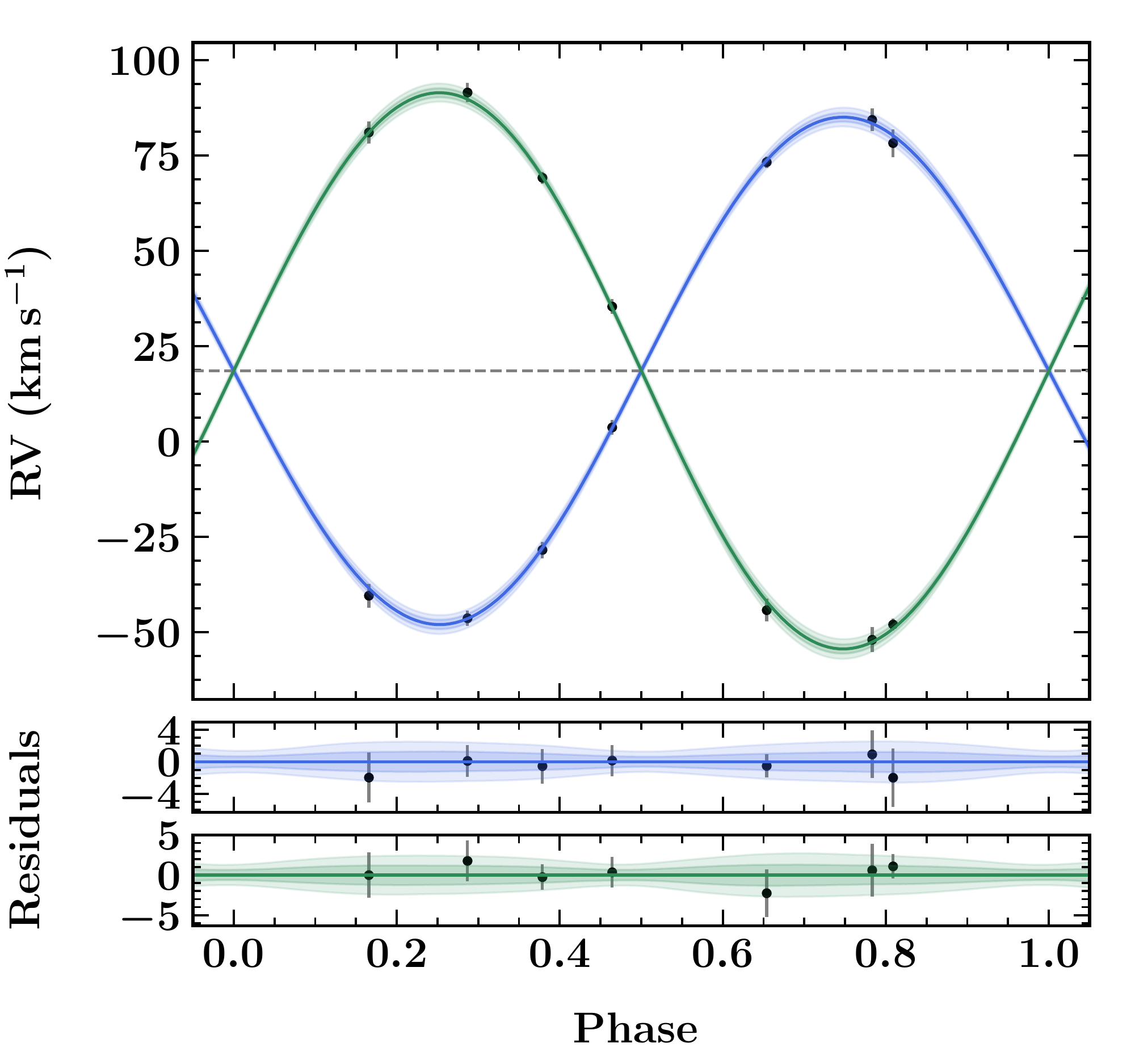}
    \caption{Phase-folded RV orbit of \name. The lines and shaded regions indicate the median and 1 \& 2$\sigma$ confidence intervals of the posterior RV orbits (blue and green for the primary and secondary stars, respectively), with Keck/HIRES RV measurements shown as black points. The grey horizontal dotted line indicates the systemic velocity. The two panels below show the residuals of the primary and secondary RV model orbits (top and bottom, respectively). The primary star's orbit has a slightly smaller semi-amplitude than the secondary's, suggesting a slightly higher primary star mass.}
   \label{fig:orbit}
\end{figure}

\begin{figure}
   \centering
   \includegraphics[width=\linewidth]{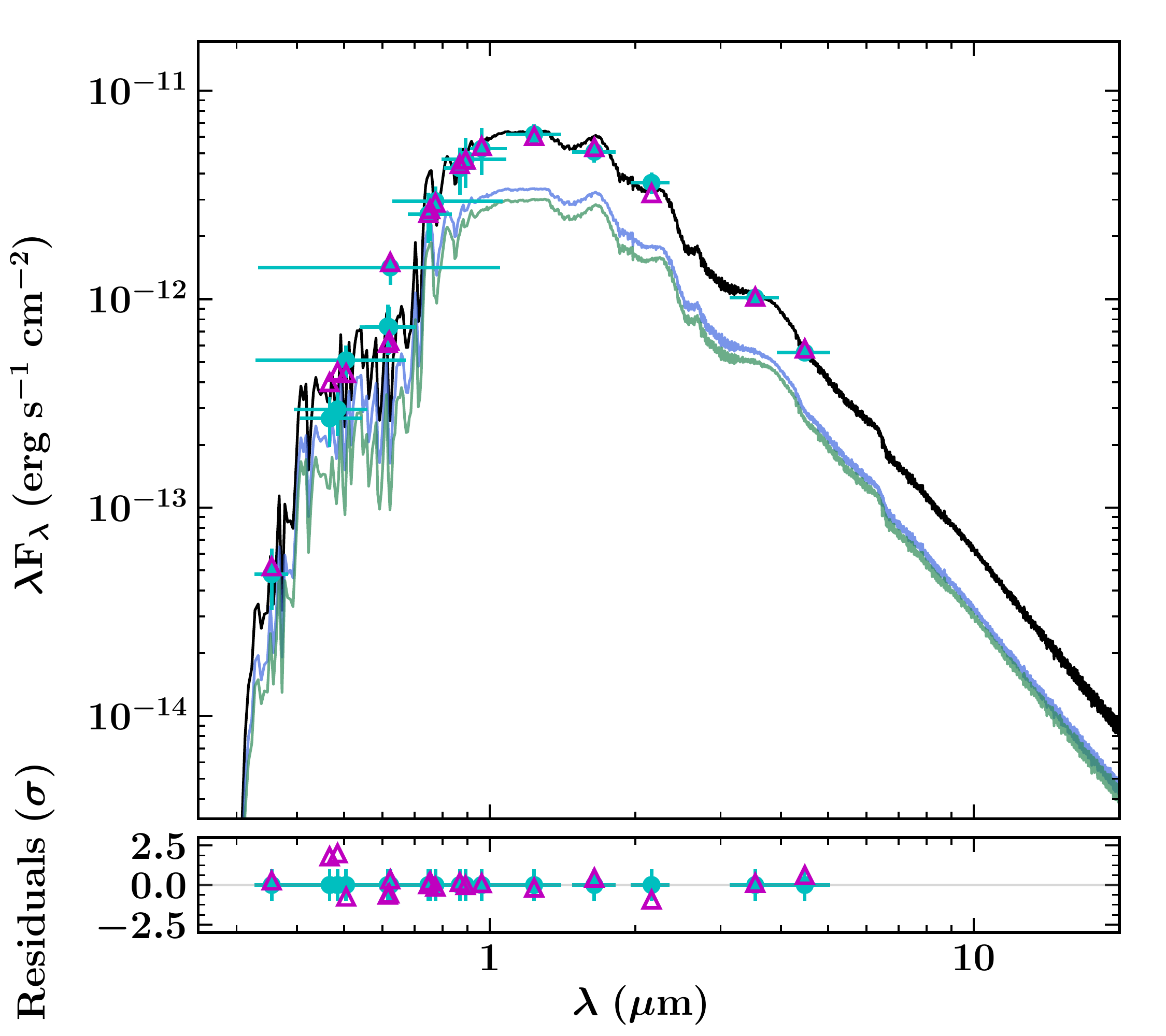}
    \caption{Spectral energy distribution (SED) of \name. Cyan points represent the observed broadband magnitudes reported in Table \ref{tab:photometry}, which together comprise the observed SED. The horizontal cyan error bars indicate the spectral range of each band. SEDs constructed from \btsettl\ model atmospheres for the primary and secondary stars are shown in blue and green, respectively. Their combined SED is shown in black with its prediction in each observed band indicated by magenta triangles. Residuals are shown below. The radii and effective temperatures of the two stars are similar, which results in comparable stellar SEDs.}
   \label{fig:SED}
\end{figure}

\begin{figure}
   \centering
   \includegraphics[width=\linewidth]{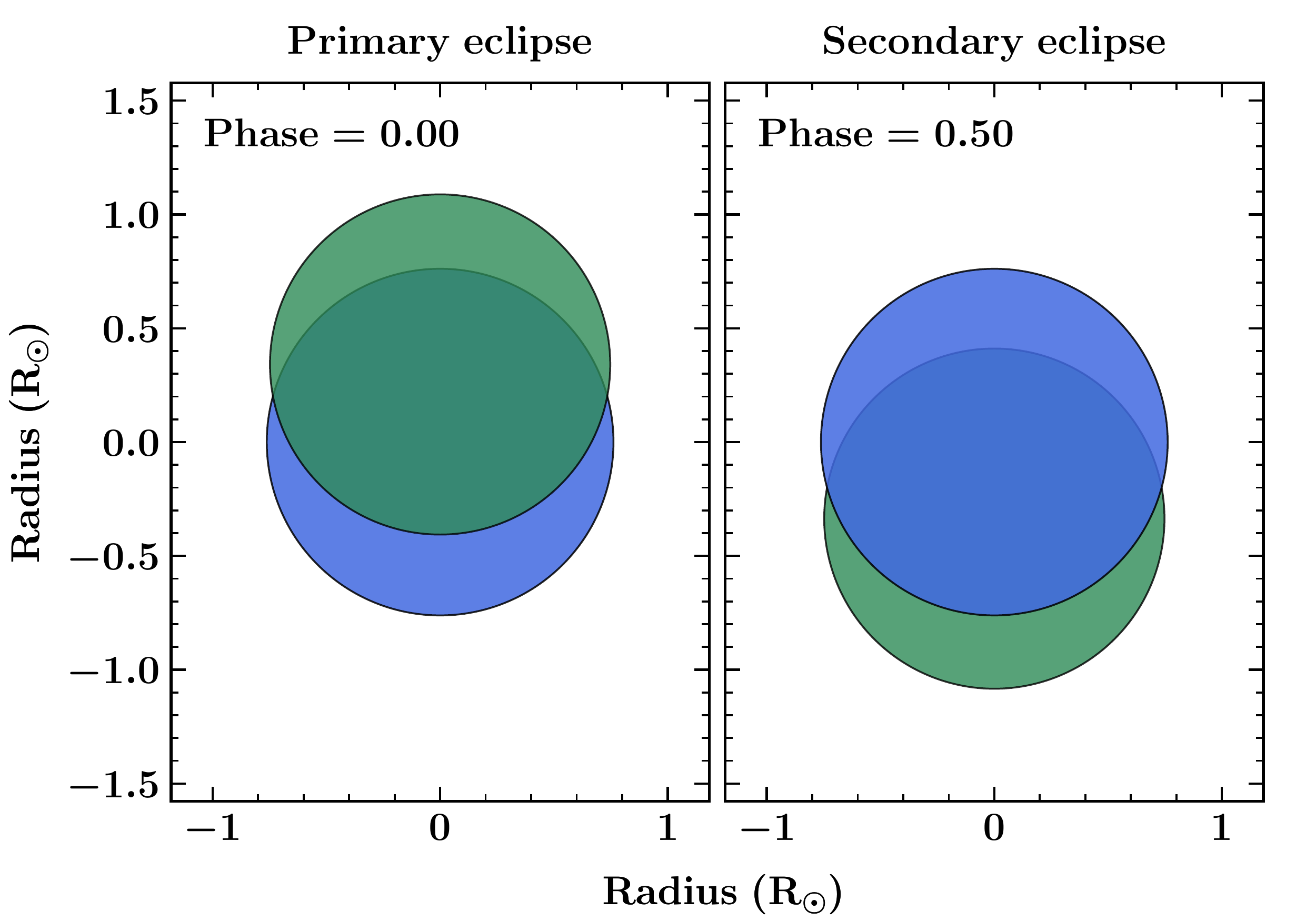}
    \caption{System geometry of \name, to scale, as observed at primary and secondary eclipse (left and right, respectively). In both cases, the primary star is shown in blue and the secondary in green. The fraction of each star occulted during eclipse is significant, which results in deep eclipses, yet they are still grazing due to the similar sizes of the two stars.}
   \label{fig:geometry}
\end{figure}

\gpe\ is an eclipsing binary and transiting planet model that is optimised for modelling young and/or active systems. Its main motivation is to account for the effect of stellar activity in the determination of stellar or planetary fundamental parameters. It is introduced in detail in \citet[][hereafter \colblue{G17}]{Gillen17a} and we refer the interested reader there. Here, we briefly recap the main points of the model and highlight some new improvements implemented since \colblue{G17}.

\gpe\ comprises a Gaussian process (GP) framework, which contains an EB and transiting planet model, and has an MCMC (Markov Chain Monte Carlo) wrapper. The central EB model uses `eb' \citep{Irwin11,Irwin18}, which is a descendent of the EBOP family of models, but which uses the analytic method of \citet{Mandel02} to perform the eclipse calculations. \gpe\ further parameterises the eccentricity and longitude of periastron terms using their combinations $\sqrt{e} \cos \omega$ and $\sqrt{e} \sin \omega$, rather than $e \cos \omega$ and $e \sin \omega$, to avoid inducing a prior on the orbital eccentricity proportional to $e^{2}$ \citep{Ford06,Triaud11}. Limb darkening is typically parameterised using the triangular sampling method of \citet{Kipping13}, with theoretical constraints applied based on the predictions of the Limb Darkening Toolkit (LDtk; \citealt{Parviainen15}) for optical and near-infrared bandpasses. The MCMC model uses the affine invariant method implemented in \textsf{emcee} \citep{Foreman-Mackey13}.

There are two main updates to \gpe\ since G17:
\begin{enumerate}
\item The GP model is now updated to use the \textsf{celerite} library \citep{Foreman-Mackey17}, which allows fast and scalable GP regression in one dimension.
\item \gpe\ now has the capability to simultaneously model spectral energy distributions (SEDs) along with light curves and radial velocities. 
\end{enumerate}

For the present analysis, we simultaneously modelled the observed light curves, RVs, and SED of \name. 
Jointly modelling the SED allows us to propagate our optical spectroscopic light ratio constraints into the IR \spitzer\ light curve bands and use them to help constrain the fundamental parameters of both stars. This is particularly important for \name, as the system displays near-equal eclipse depths in the \spitzer\ light curves, which can lead to a degeneracy between the radius ratio, surface brightness ratio and inclination when modelling the data.
Details of the modelling are given in \S \ref{sec:LCs}--\ref{sec:SED} below.

\subsection{Light curves}
\label{sec:LCs}

The \spitzer\ 3.6 and 4.5 \um\ light curves are shown in Figure \ref{fig:LCs}. In addition to eclipses on both the primary and secondary stars, both light curves display reasonably rough aperiodic out-of-eclipse (OOE) variations, which are similar to \spitzer\ light curves of other young EBs \citep[e.g.][]{Morales-Calderon12,Gillen17}. Given the rough OOE variations, we opted to use a Matern kernel with smoothness parameter $\nu$=3/2 (as approximated within the \textsf{celerite} package), which is appropriate for such rough variations \citep{Rasmussen06}. The Matern-3/2 covariance kernel $k$ is given by
\begin{equation}
k_{\rm M32} (\tau)    =   A^{2} ~ \left( 1 + \frac{\sqrt{3}\tau}{l} \right) ~ \exp \left( -\frac{\sqrt{3}\tau}{l} \right)
\end{equation}
\noindent
where $\tau = |t_{i}-t_{j}|$ is the time interval between times $t_{i}$ and $t_{j}$, and $A$ and $l$ are the characteristic amplitude and timescale of the variations, respectively.

We applied theoretical constraints on the quadratic limb darkening parameters using the predictions of \citet{Claret12}. 
We used the Claret tables rather than LDtk (Limb Darkening toolkit) because the specific intensities of the PHOENIX model spectra \citep{Husser13}, which LDtk is based on, only extend up to a wavelength of 2.6 \um, so the IRAC 3.6 and 4.5 \um\ bandpasses are not covered. We assumed \teff\ = 3200 K and \logg\ = 4.0 for both the primary and secondary stars (setting Z=0.0, which is the closest value available to the cluster metallicity; \citet{King00}\footnote{\cluster\ is suggested to have [Fe/H]\,=\,-0.15 with near-solar ratios of other elements.}). The \teff\ and \logg\ values were set based on preliminary fits to the SED data (using PHOENIX models) and the light curves and radial velocities, respectively, and then rounded to the nearest value in the Claret tables. Uncertainties on the limb darkening coefficients (LDCs) were set based on the spread of coefficient values within a representative range in \teff\ and \logg, which were determined from initial modelling of the light curves and RVs, and further inflated by comparing the Claret LDC values using both the least-square and flux conservation methods. In practice, given the IR bandpasses and grazing eclipses, the exact choice of LDCs does not significantly affect the fundamental stellar parameters.

\begin{table*}  
 \centering  
 \caption[Model parameters]{Fitted and derived parameters for \name.}  
 \label{tab:params}  
 \begin{tabular}{l l c l }  
 \noalign{\smallskip} \hline  \hline \noalign{\smallskip}  
 Parameter  &   Symbol  &  Value  &  Unit \\   
 \noalign{\smallskip} \hline 
\multicolumn{4}{c}{\emph{Fitted physical parameters}} \\   
\hline  \noalign{\smallskip}

 Orbital period    &    $P$    &  $1.9751388\pm0.0000050$   &    days      \\
 Time of primary eclipse centre    &    $T_{\rm{prim}}$    &    $2456408.9725\pm0.0013$   &  HJD \\
 Sum of radii    &    $(R_{\rm{pri}} + R_{\rm{sec}})/ a$    &    $0.2767\pm0.0017$   &   \\
 Radius ratio    &    $R_{\rm{sec}} / R_{\rm{pri}}$    &  $0.981\pm0.054$   &   \\
 Cosine of orbital inclination & $\cos i$ &    $0.0621\pm0.0015$   &     \\
 Eccentricity and argument of-    &   $\sqrt{e} \cos \omega$   &  $0.0006\pm0.0048$   &  \\
 -periastron combination terms    &   $\sqrt{e} \sin \omega$   &  $-0.083\,^{+0.072}_{-0.037}$   &   \\
 
 Systemic velocity    &    $V_{\rm{sys}}$    &    $18.47\pm0.60$   &  km\,s$^{-1}$  \\
 Primary RV semi-amplitude    &    $K_{\rm{pri}}$    &   $66.5\pm1.1$   &  km\,s$^{-1}$   \\
 Secondary RV semi-amplitude    &    $K_{\rm{sec}}$    &   $72.9\pm1.1$   &  km\,s$^{-1}$   \\
 
 Distance    &    $d$    &  $718\pm27$   &  pc    \\
 Reddening    &    A$_{\rm V}$  &  $0.105\,^{+0.150}_{-0.078}$  &      \\
 Primary effective temperature    &    $T_{\rm pri}$    &  $3260\,^{+73}_{-63}$   &  K   \\
 Secondary effective temperature    &    $T_{\rm sec}$    &  $3213\,^{+73}_{-64}$   &  K  \\

 
\noalign{\smallskip} \hline
\multicolumn{4}{c}{\emph{Derived fundamental parameters}} \\  
\hline \noalign{\smallskip}

 Primary mass      &    $M_{\rm pri}$    &  $0.2918\pm0.0099$  &    M$_{\odot}$  \\
 Secondary mass    &    $M_{\rm sec}$    &  $0.2661\,^{+0.0095}_{-0.0089}$  &    M$_{\odot}$  \\
 Primary radius    &    $R_{\rm pri}$    &  $0.762\pm0.022$  &    R$_{\odot}$  \\
 Secondary radius  &    $R_{\rm sec}$    &  $0.748\pm0.023$  &    R$_{\odot}$  \\
 Primary luminosity    &    $L_{\rm pri}$    &  $0.0590\,^{+0.0064}_{-0.0052}$  &    L$_{\odot}$  \\
 Secondary luminosity    &    $L_{\rm sec}$    &  $0.0536\,^{+0.0056}_{-0.0048}$  &    L$_{\odot}$  \\
 
 Primary surface gravity    &    $\log g_{\rm pri}$    &  $4.140\pm0.025$  &    (cm\,s$^{-2}$)  \\
 Secondary surface gravity    &    $\log g_{\rm sec}$    &  $4.116\pm0.026$  &    (cm\,s$^{-2}$)  \\
 
 Mass sum     &    $M_{\rm pri} + M_{\rm sec}$    &  $0.558\pm0.019$  &    M$_{\odot}$  \\
 Radius sum    &    $R_{\rm pri} + R_{\rm sec}$    &  $1.508\pm0.019$  &    R$_{\odot}$  \\
 
 
\noalign{\smallskip} \hline
\multicolumn{4}{c}{\emph{Derived radiative, orbital and rotational parameters}} \\  
\hline \noalign{\smallskip}

 Central surface brightness ratio in \Sone    &    $J_{\rm{\Sone}}$    &    $0.965\pm0.022$     &   \\
 Central surface brightness ratio in \Stwo    &    $J_{\rm{\Stwo}}$    &     $0.971\pm0.018$    &   \\

 Semi-major axis    &    $a$    &    $5.453\pm0.060$  &  R$_{\odot}$  \\
 Orbital inclination  &  $i$  &  $86.441\,^{+0.088}_{-0.079}$  &  $^{\circ}$  \\
 Eccentricity    &    $e$    &  $0.0068\,^{+0.0075}_{-0.0062}$  &    \\
 Longitude of periastron    &    $\omega$    &  $-89\,^{+13}_{-3}$  &    $^{\circ}$  \\

 Primary synchronised velocity      &    $V_{\rm pri ~ sync}$    &  $19.51\pm0.57$  &  km\,s$^{-1}$  \\
 Secondary synchronised velocity    &    $V_{\rm sec ~ sync}$    &  $19.15\pm0.59$  &  km\,s$^{-1}$  \\

 \noalign{\smallskip}  
 \hline  
 \end{tabular}
 \end{table*}

\subsection{Radial Velocities}
\label{sec:RVs}

For each Keck/HIRES spectrum, spectral orders between 6200--7200\,\AA\ were used to determine RVs for both components of \name\ via cross-correlation. In each analysed order, the spectral range was restricted to avoid the edges of orders and telluric lines. The task $fxcor$ within IRAF was used to measure relative velocities between \name\ and a set of $\sim$M2--M4 radial velocity reference stars (i.e. well-matched to our M3 source).  The number of reference stars ranged between 2 and 14, depending on the observation date, with  GJ 105B, GJ 388, GJ 176, and GJ 411 the most typically sampled stars over the many nights and runs.  Correlation peaks were measured using either Gaussian or parabolic fits, depending on the signal-to-noise (S/N). The resulting RV time-series was modelled with Keplerian orbits. 

The correlation peaks were further used to establish the red optical flux ratio at the measurement epochs. We used the ratio of the two cross-correlation peak heights to give an estimate of the light ratio between the two stars in the 6200--7200\,\AA\ range. Spectral disentangling is an alternative approach to estimate stellar parameters for each star individually and hence is amenable to estimating flux ratios. It has been successfully applied to more massive systems \citep[see e.g.][]{Pavlovski18,Johnston19}, as well as low-mass objects through the use of GPs \citep{Czekala17}, and we therefore see it as an interesting avenue for further work.

Spectroscopic light ratios can be used as a prior during the light curve and RV fitting process. They are particularly useful for EBs with near-equal eclipse depths, as they can help break the degeneracy between the radius and surface brightness ratios, and also the inclination. Typically, such spectroscopic light ratios are used as direct priors on the light curve solution, under the assumption that the spectroscopic wavelength range is an acceptable proxy for the observed light curve bandpass. Here, however, this is not possible because we have optical spectra and IR light curves. We therefore need a way of propagating the light ratio constraint from the optical spectra into the \spitzer\ light curve bands; we do this by simultaneously modelling the SED of the system (i.e. the joint SED of both stars).

\subsection{Spectral Energy Distribution}
\label{sec:SED}

Publicly available broadband photometry (as reported in Table \ref{tab:photometry}) typically cover the near-UV to the near/mid-IR. This wavelength range covers the rise, peak and tail of photospheric emission from low-mass stars, which means that SEDs can provide useful constraints on their temperatures. Furthermore, with Gaia DR2 parallaxes, SEDs offer complementary information on the stellar radii.

We model the observed SED of \name\ as the sum of two stellar photospheres (primary and secondary stars), using \btsettl\ model atmospheres \citep{Allard12} and interpolating the model grid in \teff\,--\,\logg\ space\footnote{We also tested using the PHOENIX v2 model atmospheres \citep{Husser13} but these provided a slightly worse, although still acceptable, fit. Comparing the effect of using \btsettl\ or PHOENIX models is discussed in \S \ref{sec:BT_vs_PX}.}. Magnitudes and uncertainties (in either the AB or Vega systems) were converted to fluxes using the standard relation between magnitude and flux, with their zero point values and effective wavelengths (defined as the pivot wavelength) obtained from the relevant references in Table \ref{tab:photometry}.

The parameters of the fit were: the temperatures, radii and surface gravities of both stars, the distance and reddening to the system, and a jitter term per photometric dataset (i.e. SDSS, Gaia, 2MASS). 
The reddening model follows the extinction law of \citet{Fitzpatrick99}, which has been improved by \citet{Indebetouw05} in the IR (as available through FPS). The SED model used here is essentially an updated version of the model presented in \colblue{G17}, which can now: 
\begin{enumerate}
\item accept Gaia parallax measurements to provide distance constraints
\item apply a jitter term per photometric dataset (rather than one for the whole SED).
\end{enumerate}

By simultaneously modelling the SED, we can propagate the light ratio constraint from the optical spectra into the \spitzer\ light curve bands. We do this by applying the spectroscopic light ratio as a prior constraint on the atmospheric model light ratio between 6200--7200\,\AA, and using the corresponding model flux ratio in the \spitzer\ bands as the central surface brightness ratio for the eclipse modelling. For the spectroscopic light ratio, we use $l_{\rm sec}/l_{\rm pri}$ = 0.896\,$\pm$\,0.077, which is the median and 2$\sigma$ MAD-estimated\footnote{MAD = median absolute deviation.} standard error on the individual values given in Table \ref{tab:RVs}. In addition to enabling the use of a spectroscopic light ratio, modelling the SED allows us to solve for the stellar masses, radii and temperatures in a self-consistent manner, and to optionally take advantage of the Gaia parallax information when fitting the data.

By including stellar atmosphere models, we make the fitting process slightly model-dependent in the sense that we are relying on the stellar atmosphere models to accurately predict flux ratios between the two stars in different photometric bands. This is only a small dependence, however, because while atmosphere models may struggle to reproduce all lines and molecular features, they should be reasonably reliable for predicting the relative flux ratio between two model atmospheres in wide-field photometric bands, such as both the broadband SED photometry and the light curve bands are. We note that in \S \ref{sec:EB1039}, we show for another system (which does not significantly suffer from the aforementioned degeneracies between the radius ratio, inclination and surface brightness ratio), that modelling either the light curves and RVs only, or also modelling the SED, gives consistent masses and radii for both stars.

Finally, we note that by using stellar atmosphere models with a single temperature for each star, we are not fully accounting for the effect of starspots. Specifically, we are assuming that the overall effect of magnetically active regions to reduce the inferred \teff\ (i.e. a combination of filling factor and average spot contrast) is essentially the same on each star, so that the \teff\ ratio between the model atmospheres is an acceptable approximation of the true \teff\ ratio. For \name, where the two stars are the same age and possess similar masses and radii, it is reasonable to assume their activity levels and hence spot coverage will also be similar. Indeed, the two stars display similar strength line profiles for activity-sensitive lines in the HIRES spectra (see e.g. Figure \ref{fig:spectra}).

\section{Results}
\label{sec:results}

\begin{figure}
   \centering
   \includegraphics[width=\linewidth]{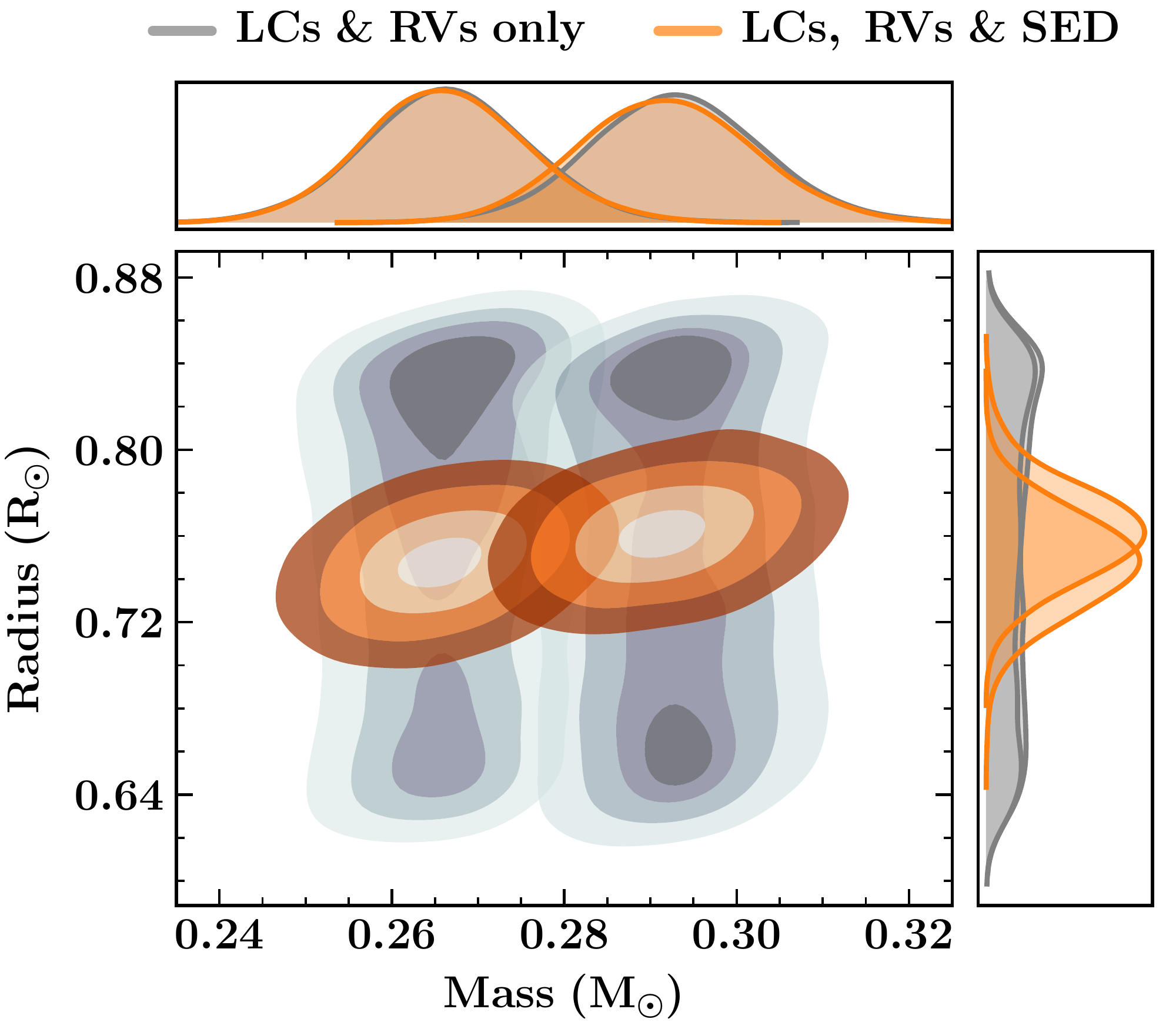}
    \caption{Assessing the effect of modelling the SED along with the light curves and RVs. Main plot: 2D posterior mass and radius distributions from modelling only the light curves and RVs (grey) compared to when the SED (and hence spectroscopic light ratio) is included in the fit (orange). The contours represent 0.5, 1, 1.5 and 2 $\sigma$ boundaries. Top and right: 1D posterior distributions for the masses and radii, respectively. Including the SED in the modelling of \name\ significantly reduces the radius uncertainties and turns the slightly bimodal radius distributions into unimodal ones. There is no significant effect on the masses, as expected.}
   \label{fig:MR_compare_SED}
\end{figure}

\begin{figure*}
   \centering
   \includegraphics[width=0.8\linewidth]{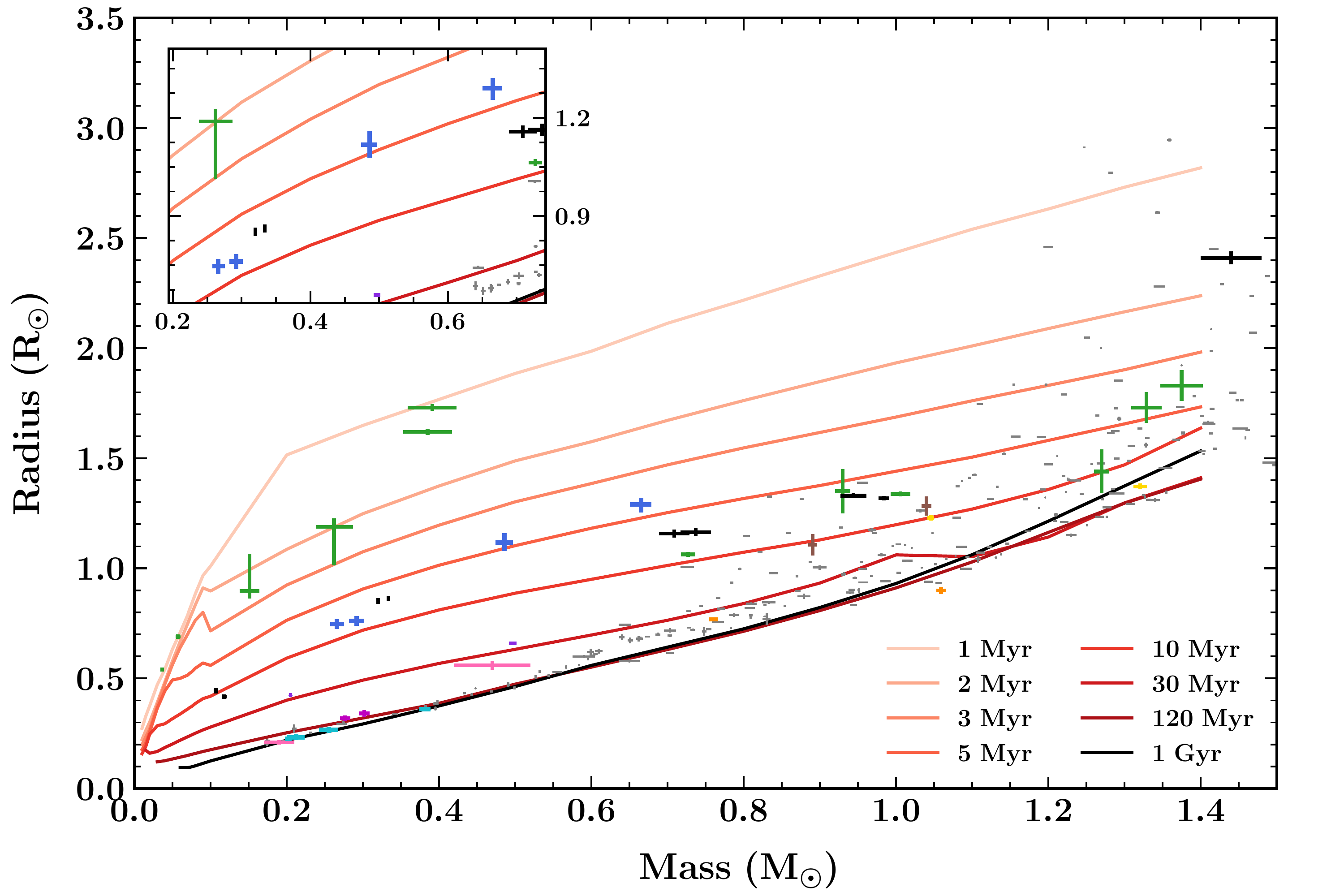}
    \caption{Mass-radius relation for detached double-lined EBs below 1.5 M$_{\odot}$. The coloured lines represent the solar metallicity isochrones of \citet[][BHAC15]{Baraffe15} from 1 Myr to 1 Gyr (light-to-dark, top-to-bottom). Well-characterised EBs in sub-Gyr open clusters are coloured while field EBs are shown in grey. The two EBs in \cluster, \name\ and \ebdisk, are plotted in blue, with other well-characterised cluster EBs shown in green (Orion), black (Upper Scorpius), pink (NGC 1647), gold (Per OB2), magenta (Pleiades), orange (Hyades), cyan (Praesepe), brown (Upper Centaurus Lupus) and purple (32 Orionis Moving Group). The cluster EBs are compiled from \citet[][Table 5]{Gillen17a}, with subsequent additions from \citet{Gomez-Maqueo-Chew19}, \citet{David19} and \citet{Murphy20}, and the field EBs from DEBCat \citep{Southworth15}. Inset: zoom on the region around \name\ and \ebdisk.}
   \label{fig:MR}
\end{figure*}


\begin{figure*}
   \centering
   \includegraphics[width=\linewidth]{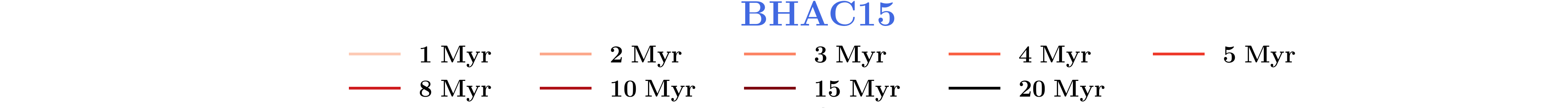}
   \includegraphics[width=\psize\linewidth]{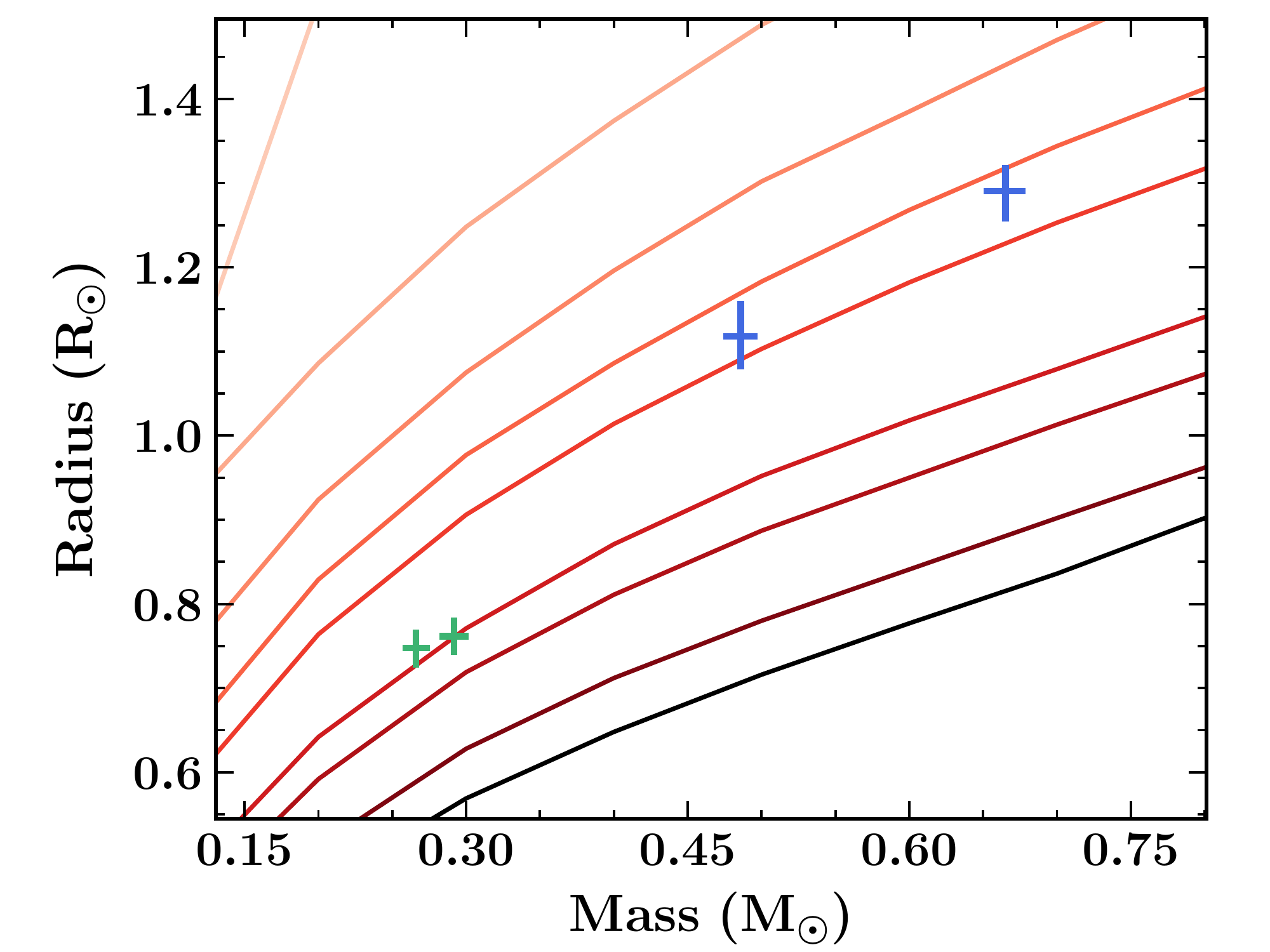}
   \includegraphics[width=\psize\linewidth]{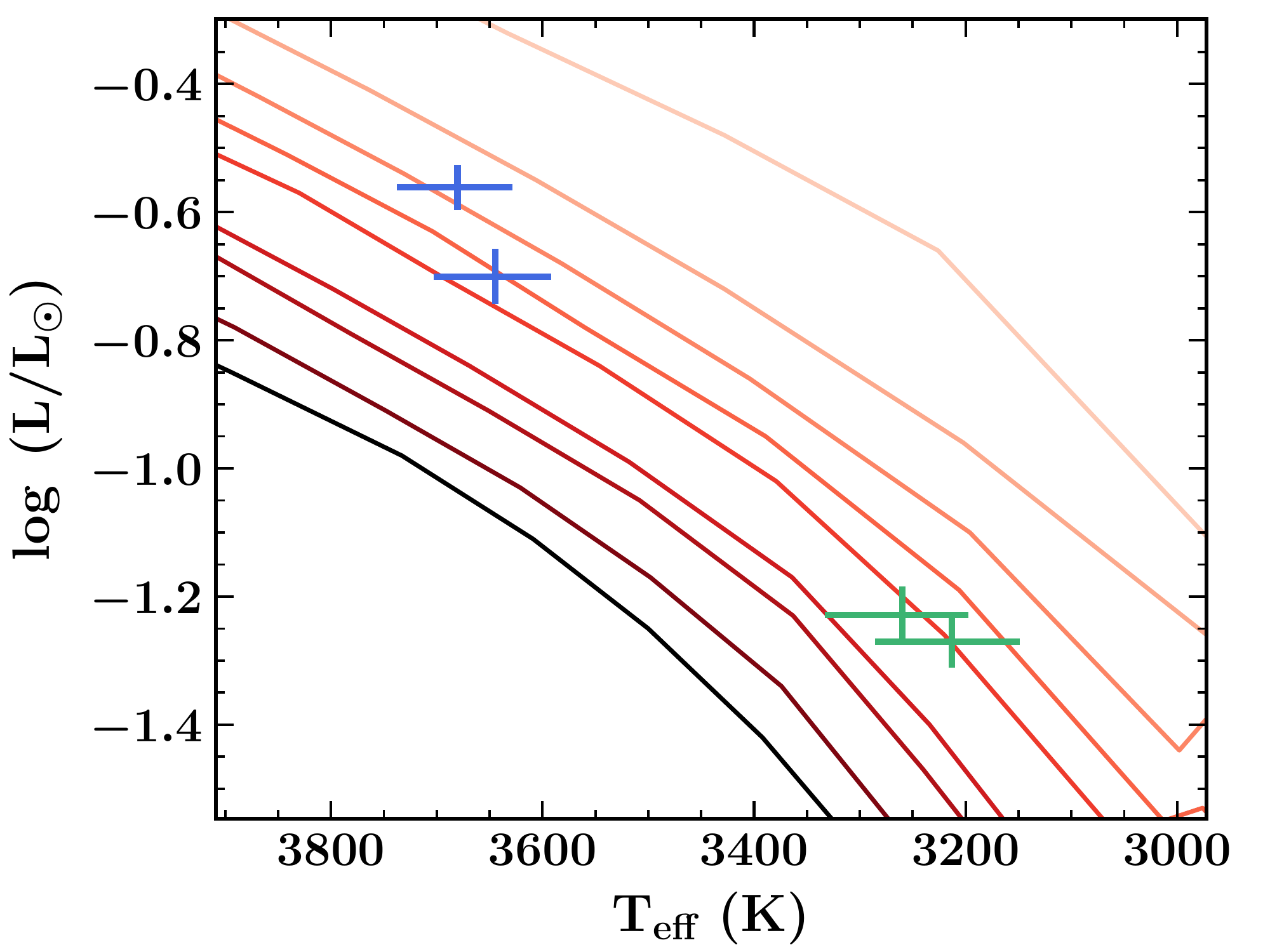} \\
   \includegraphics[width=\psize\linewidth]{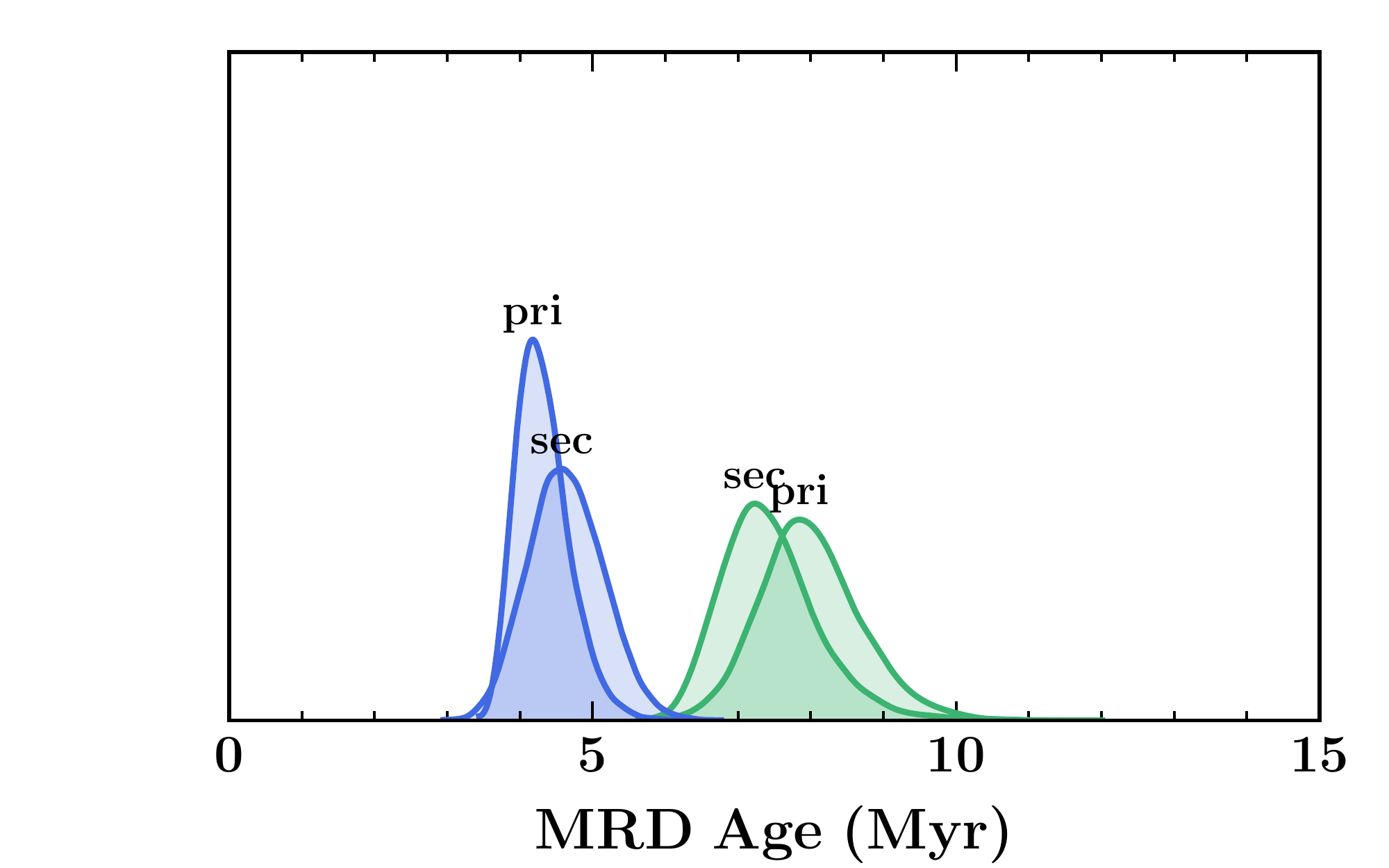}
   \includegraphics[width=\psize\linewidth]{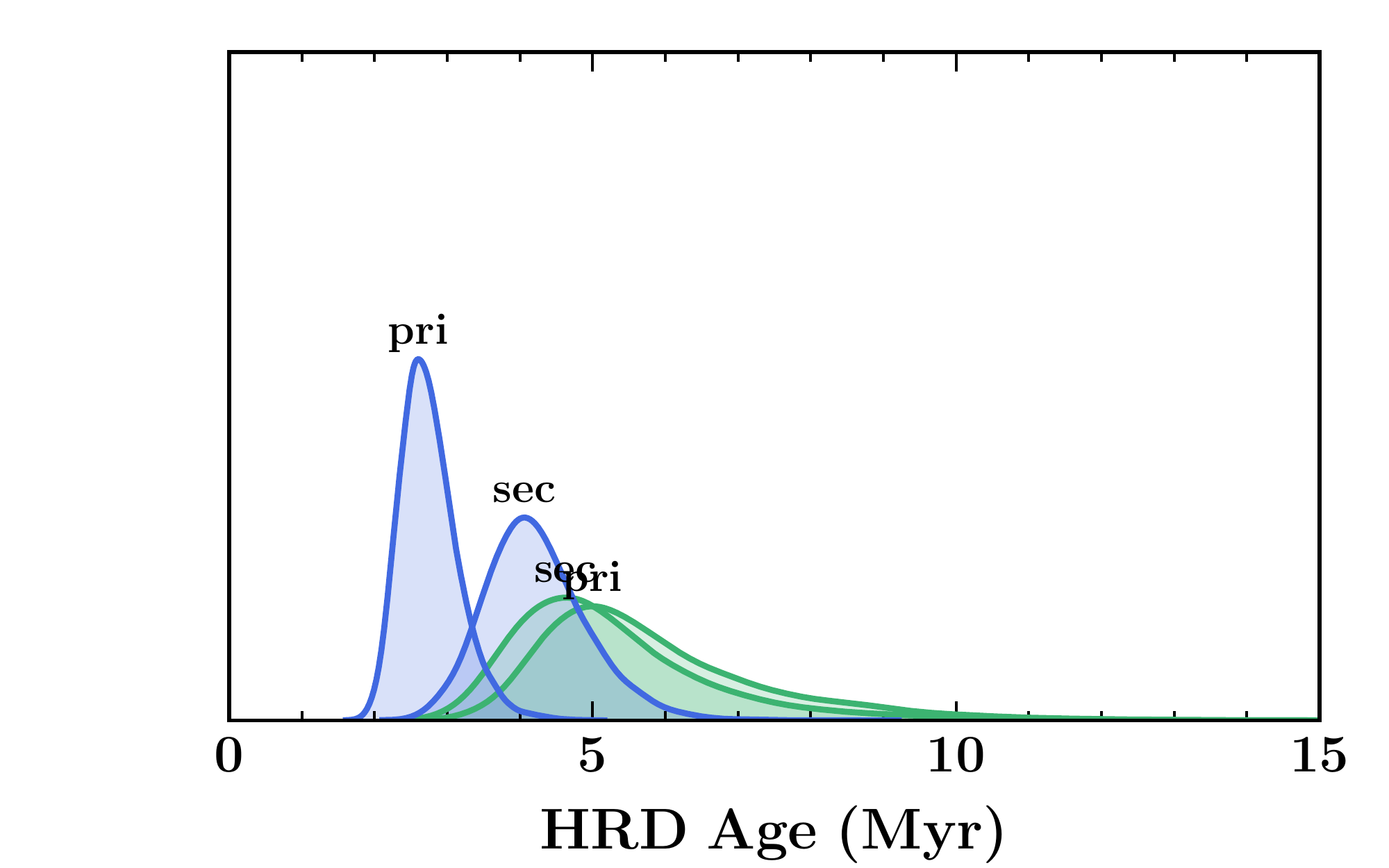} \\
   \caption{Comparison of the fundamental parameters of \name\ (green) and \ebdisk\ (blue) to the predictions of the BHAC15 stellar evolution models. \emph{Top row}: \mrp\ relation showing model isochrones between 1--20\,Myr (\emph{left}), and the \tlp\ relation with the same set of model isochrones. \emph{Bottom row}: corresponding age distributions for \name\ (green) and \ebdisk\ (blue) based on their locations in the \mrp\ diagram (MRD; \emph{left}) and \tlp\ (i.e. Hertzsprung--Russell) diagram (HRD; \emph{right}). The individual plots in Figures \ref{fig:BHAC15_compare}--\ref{fig:Parsec_compare} all share the same axes for ease of comparison.}
   \label{fig:BHAC15_compare}
\end{figure*}

\begin{figure*}
   \centering
   \includegraphics[width=\linewidth]{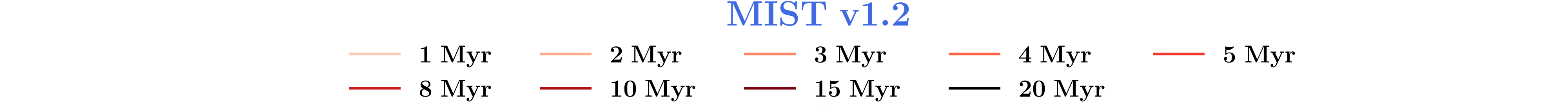}
   \includegraphics[width=\psize\linewidth]{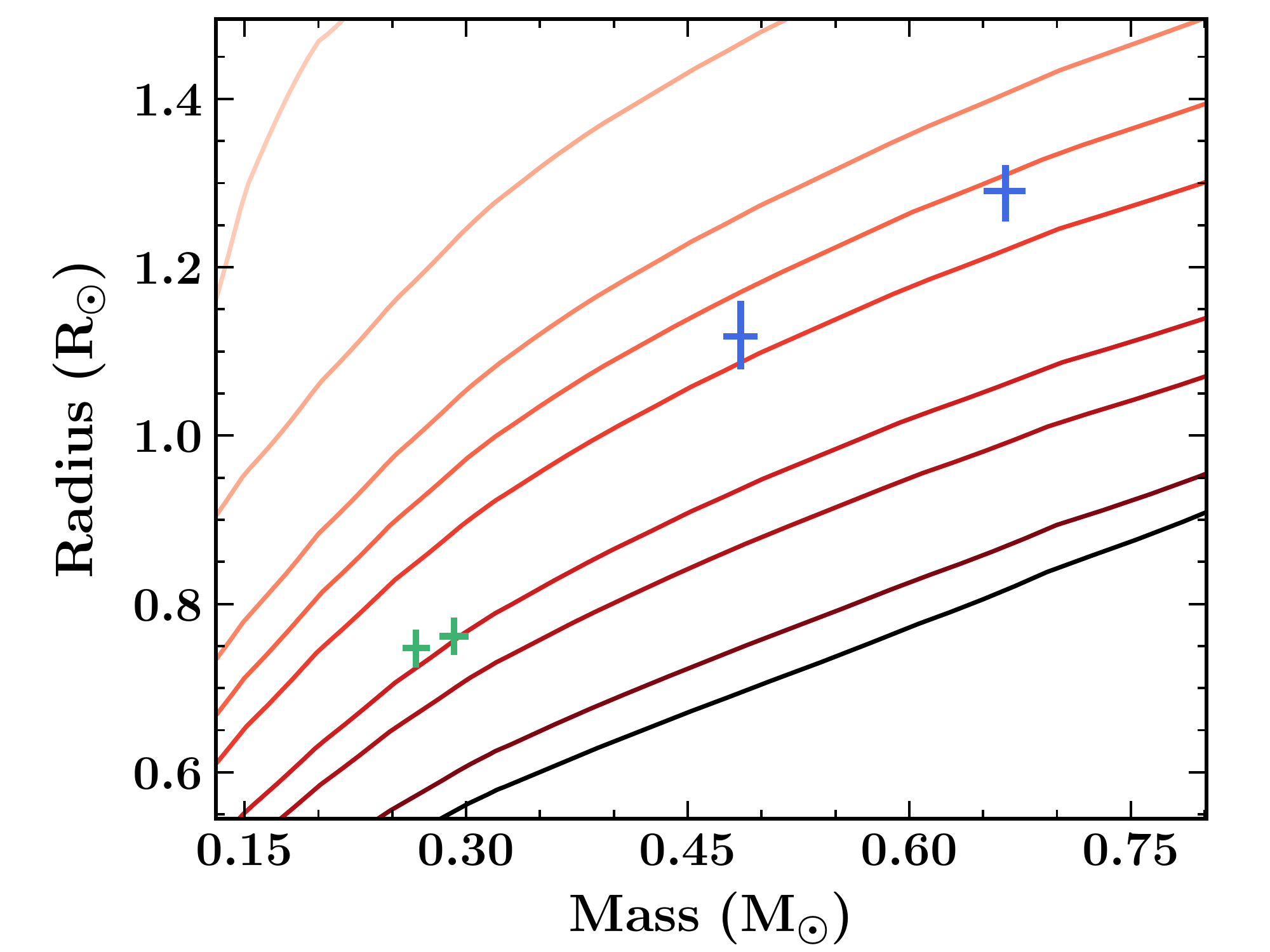}
   \includegraphics[width=\psize\linewidth]{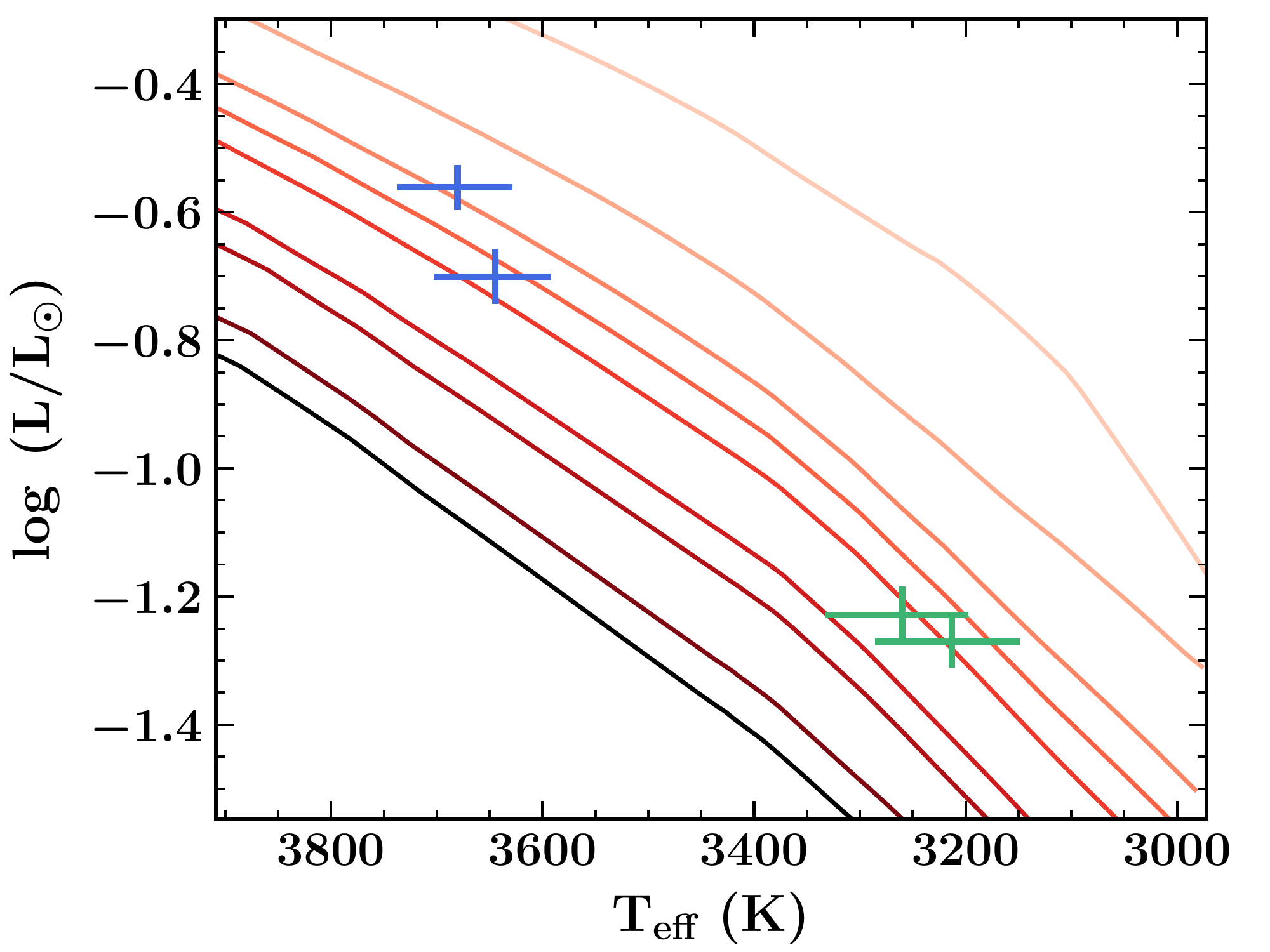} \\
   \includegraphics[width=\psize\linewidth]{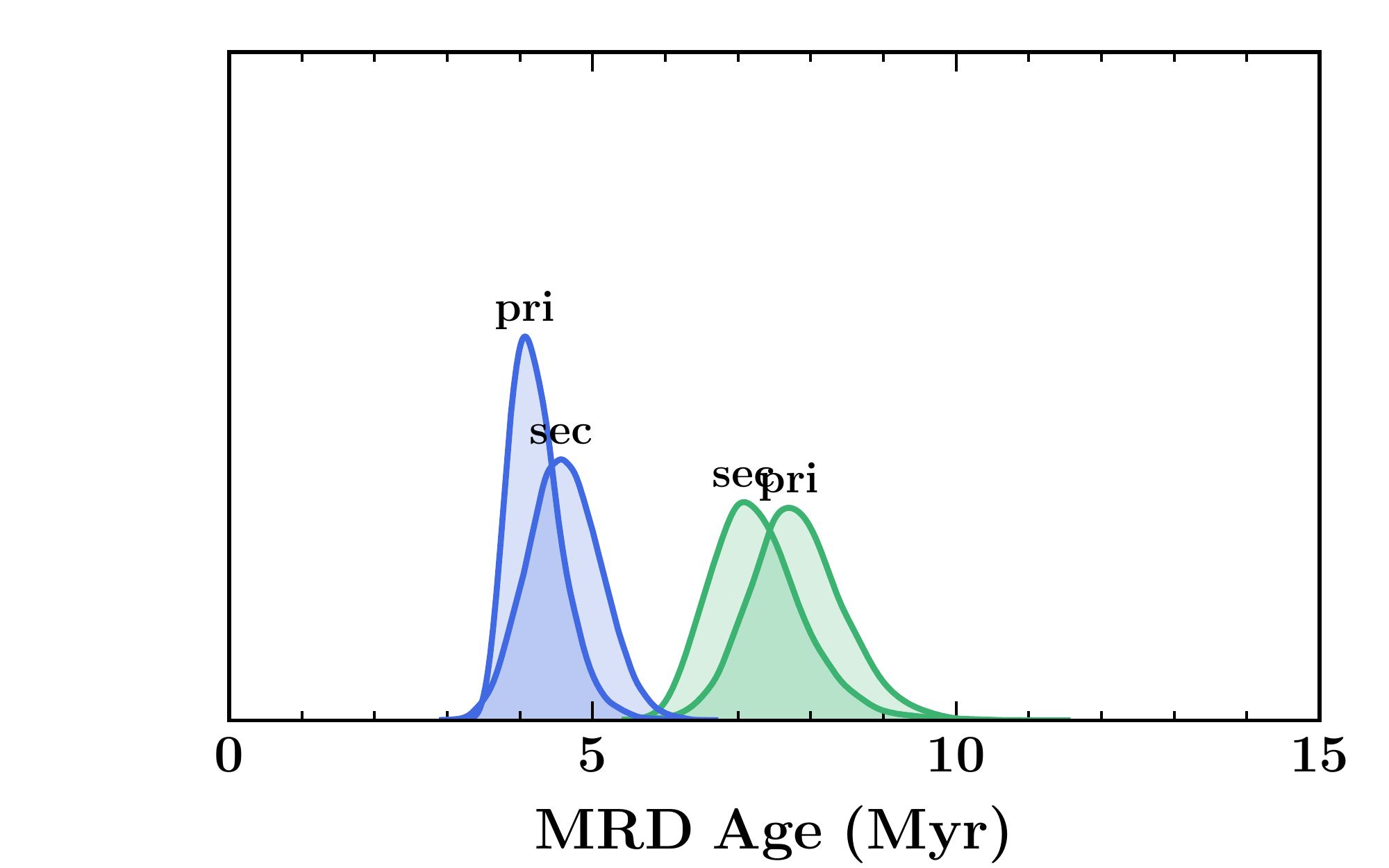}
   \includegraphics[width=\psize\linewidth]{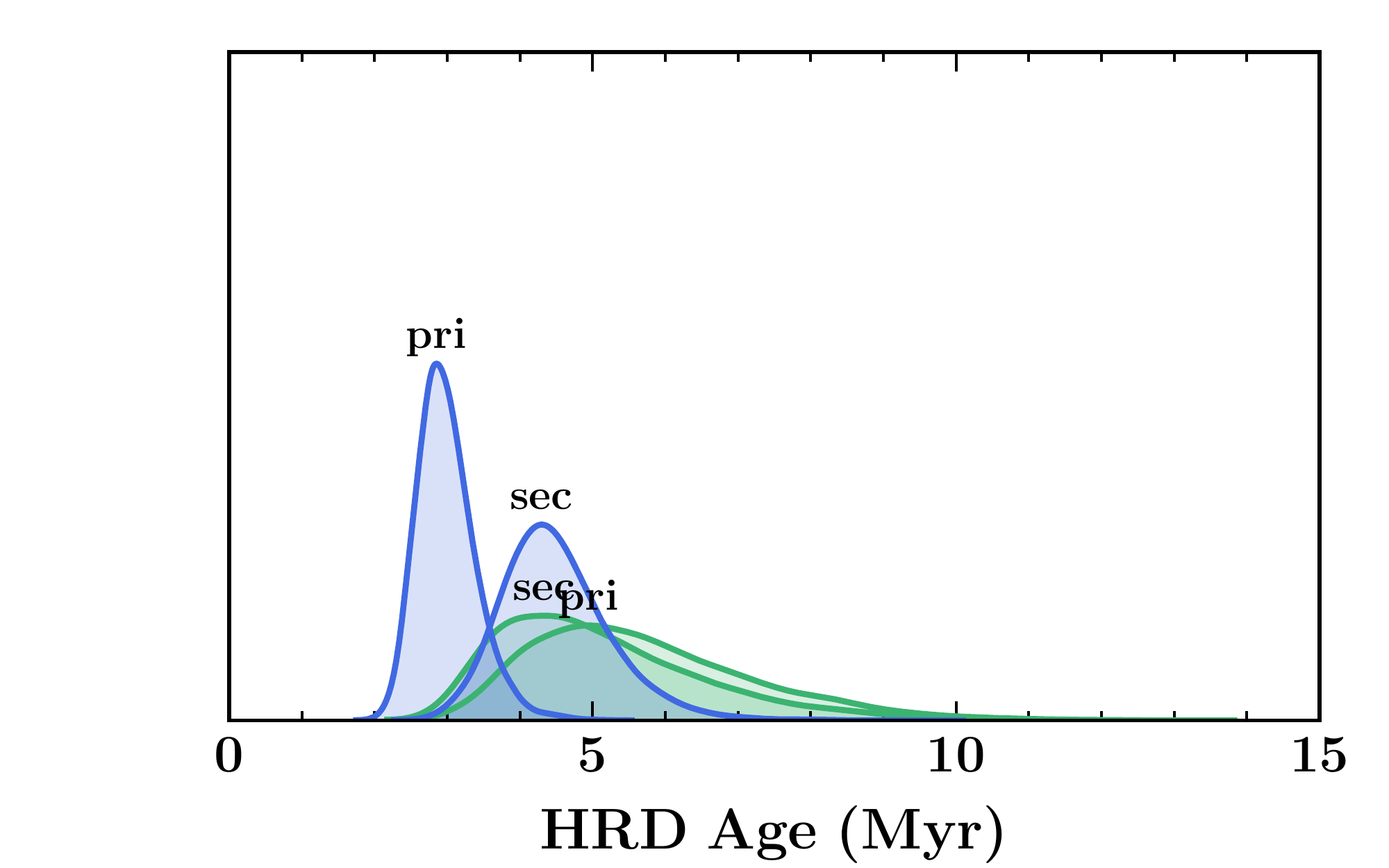} \\
   \caption{Same as Figure \ref{fig:BHAC15_compare} but comparing to MIST v1.2 models.}
   \label{fig:MIST_compare}
\end{figure*}

\begin{figure*}
   \centering
   \includegraphics[width=\linewidth]{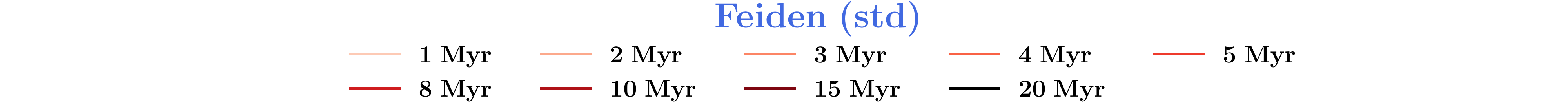}
   \includegraphics[width=\psize\linewidth]{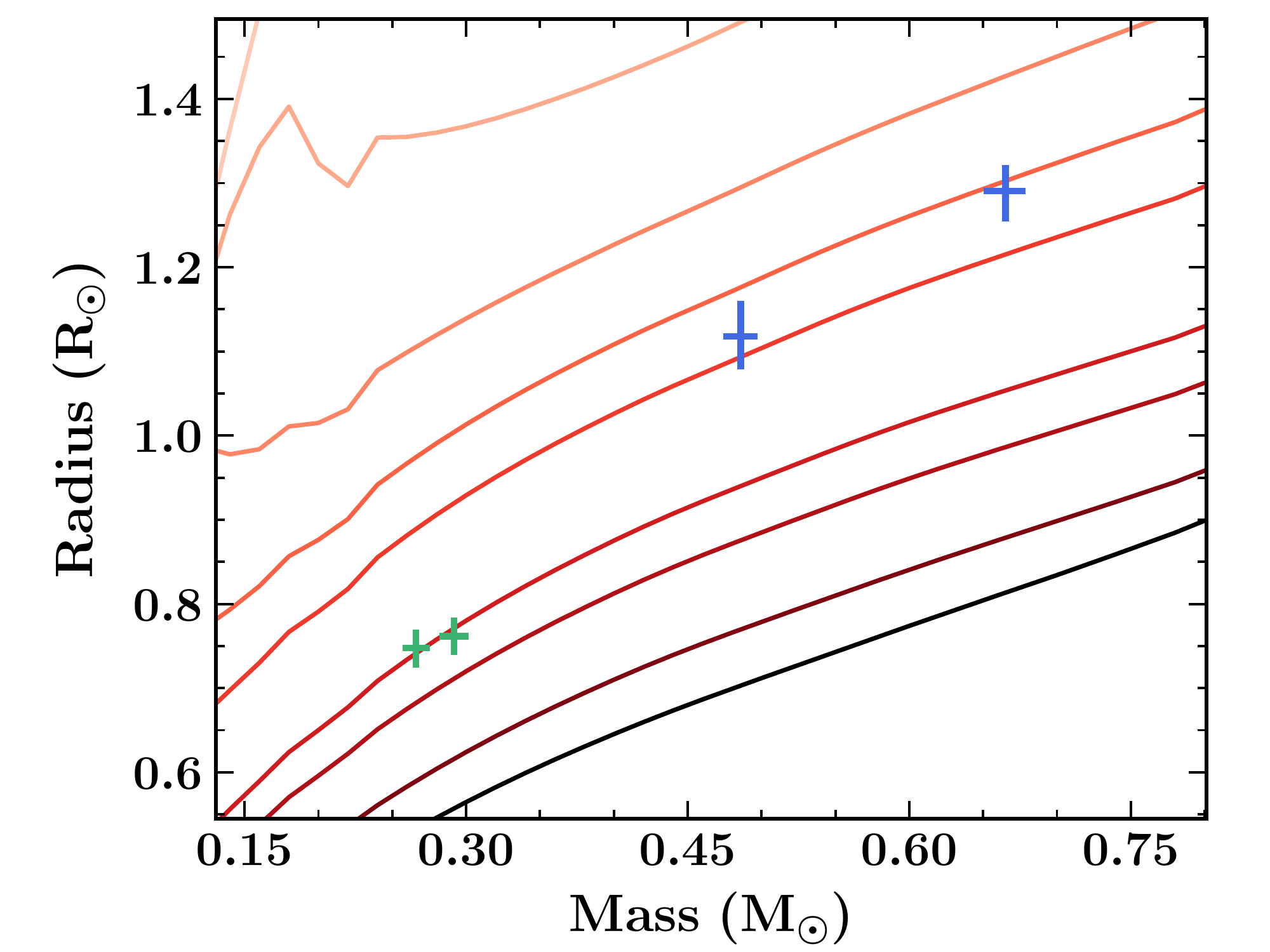}
   \includegraphics[width=\psize\linewidth]{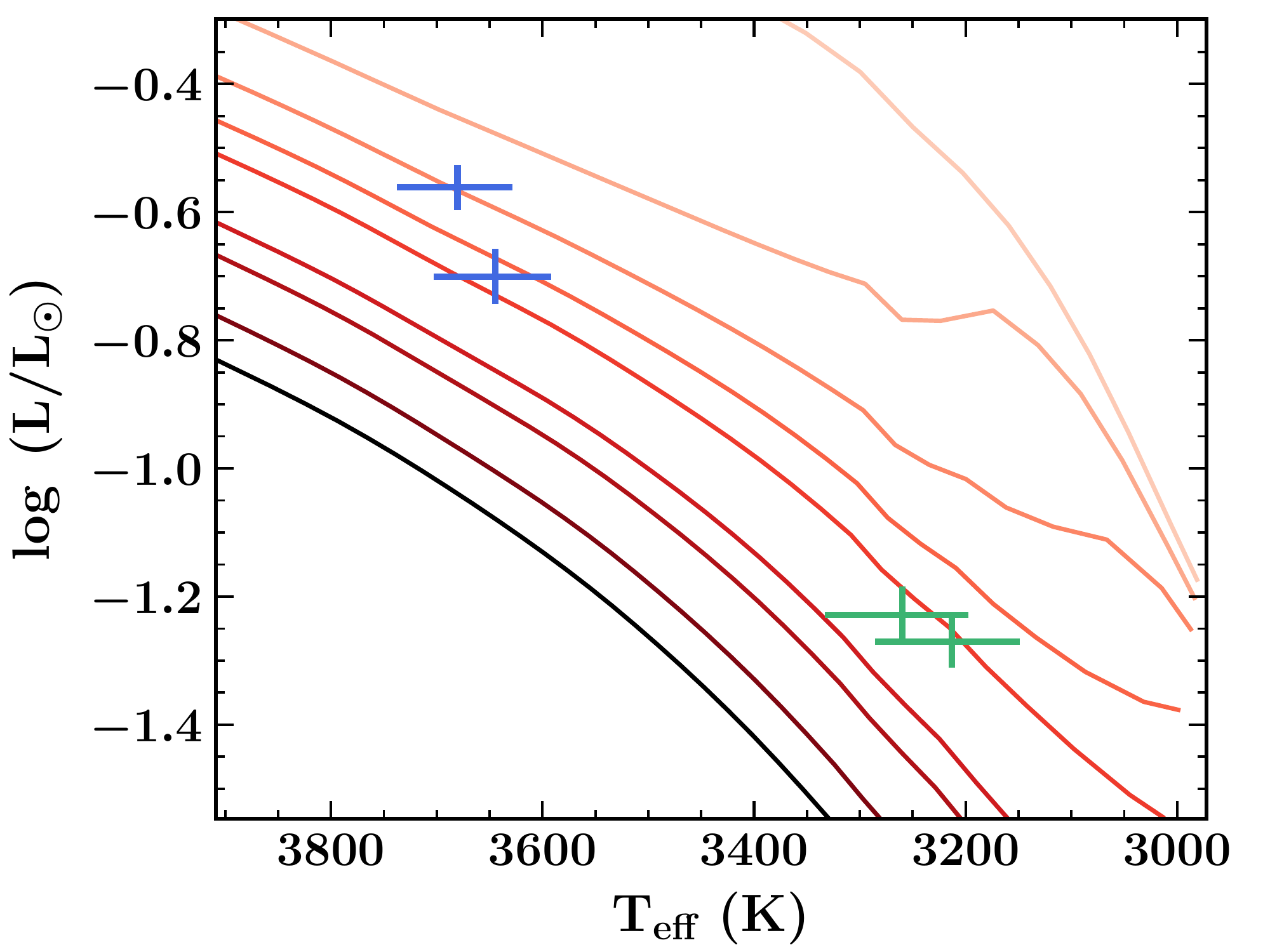} \\
   \includegraphics[width=\psize\linewidth]{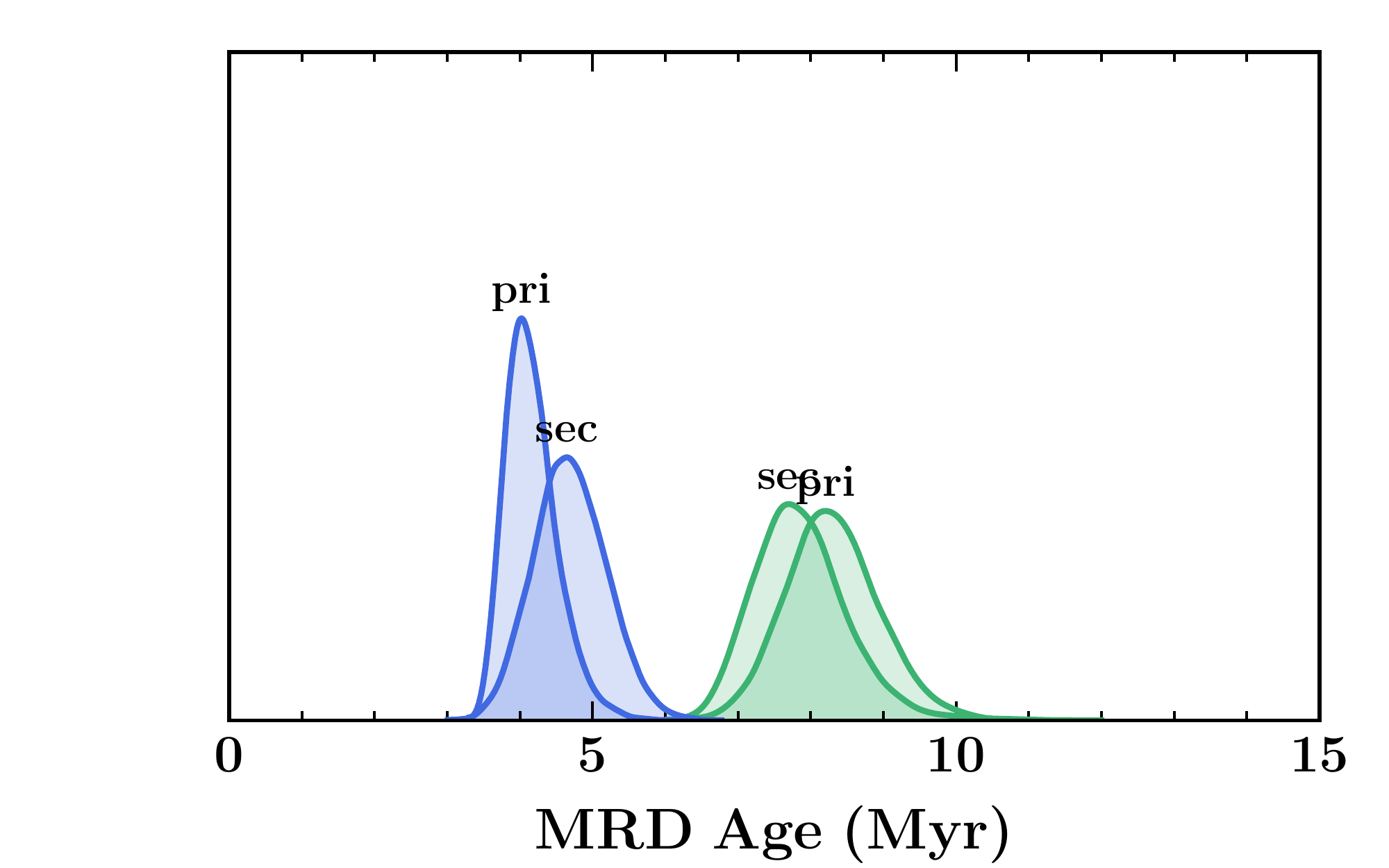}
   \includegraphics[width=\psize\linewidth]{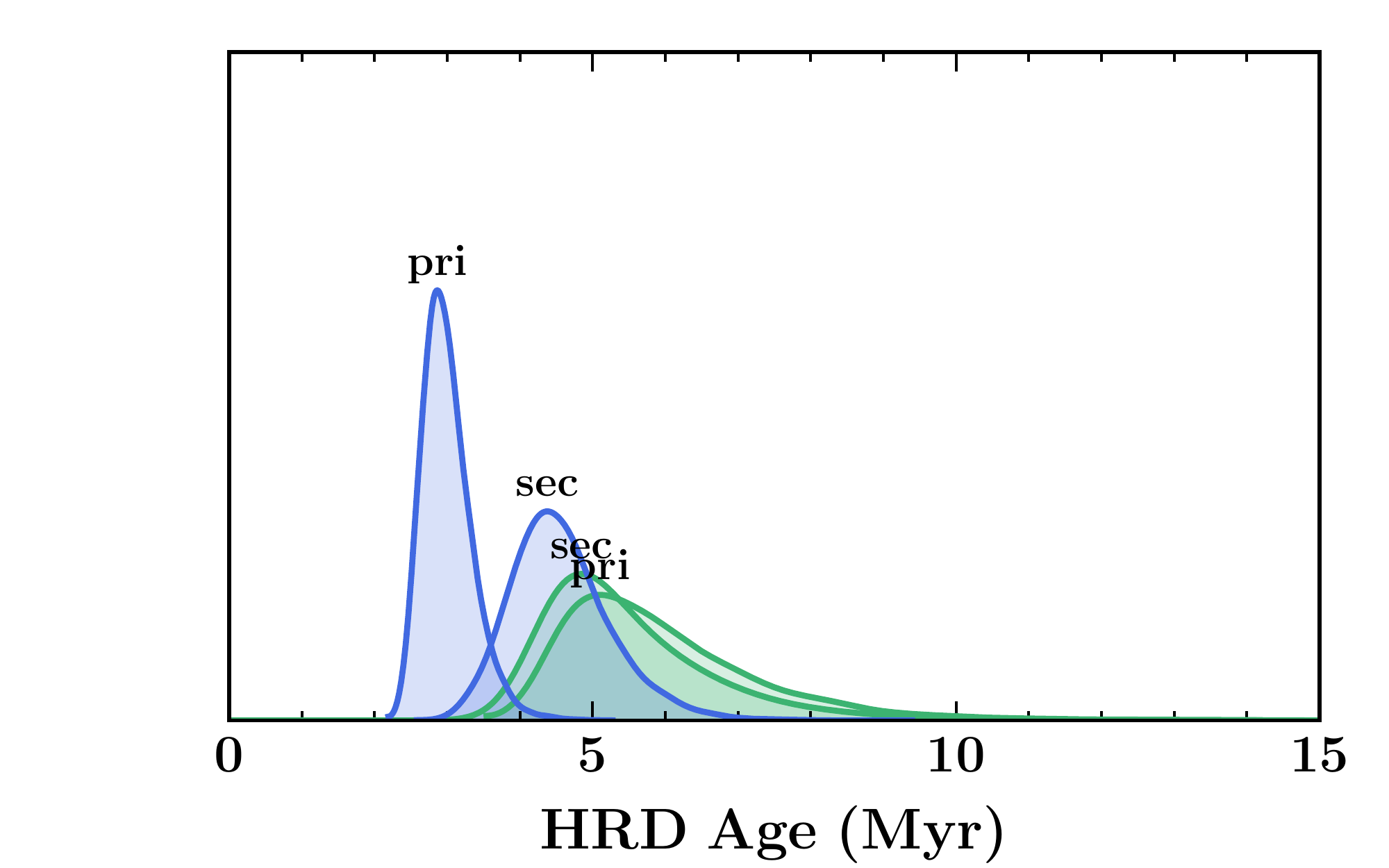} \\
   \caption{Same as Figure \ref{fig:BHAC15_compare} but comparing to the standard Feiden models.}
   \label{fig:F16std_compare}
\end{figure*}

\begin{figure*}
   \centering
   \includegraphics[width=\linewidth]{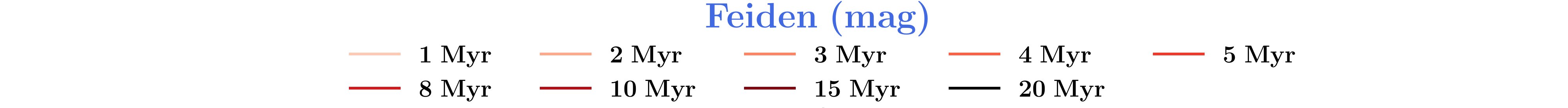}
   \includegraphics[width=\psize\linewidth]{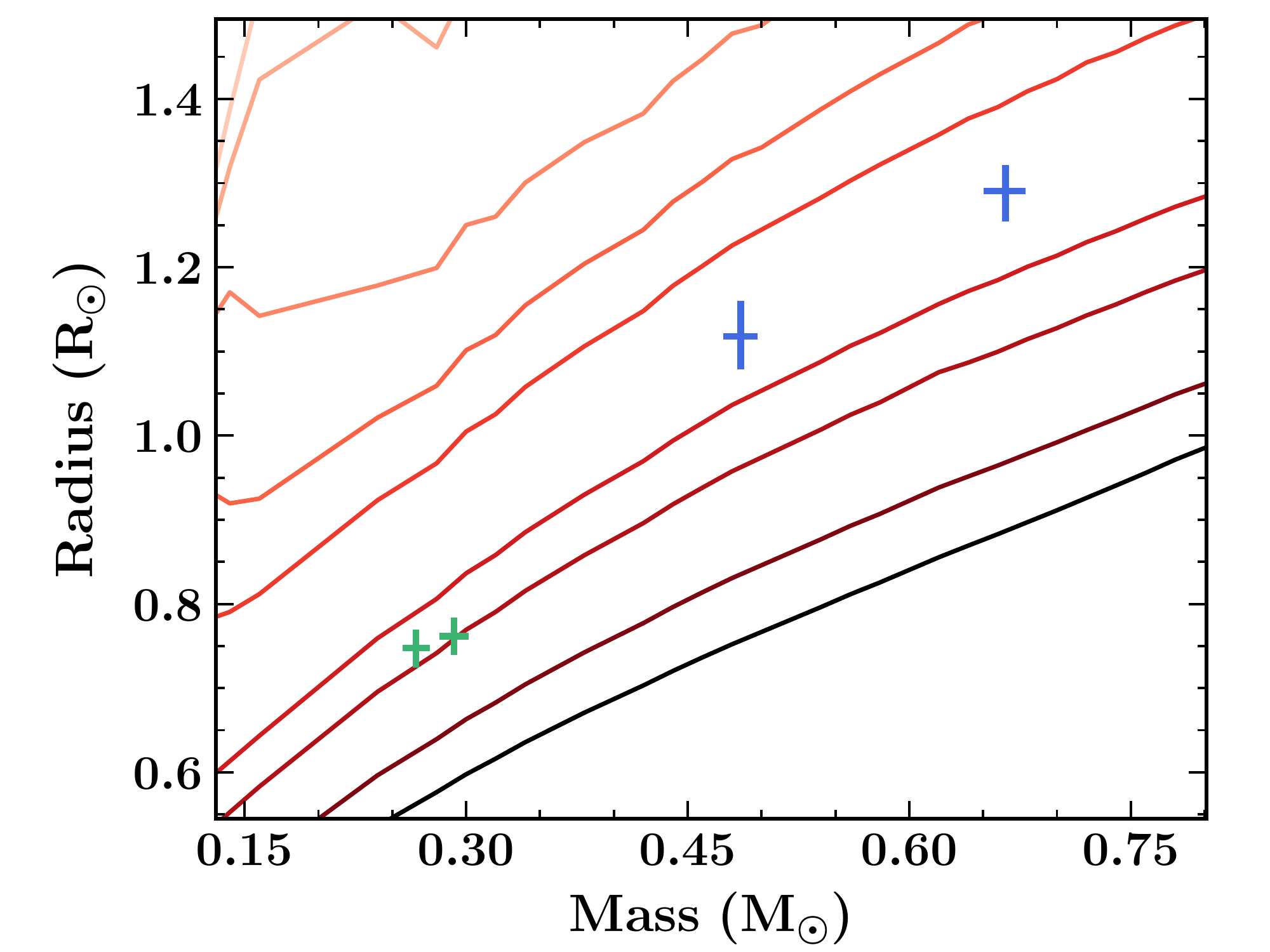}
   \includegraphics[width=\psize\linewidth]{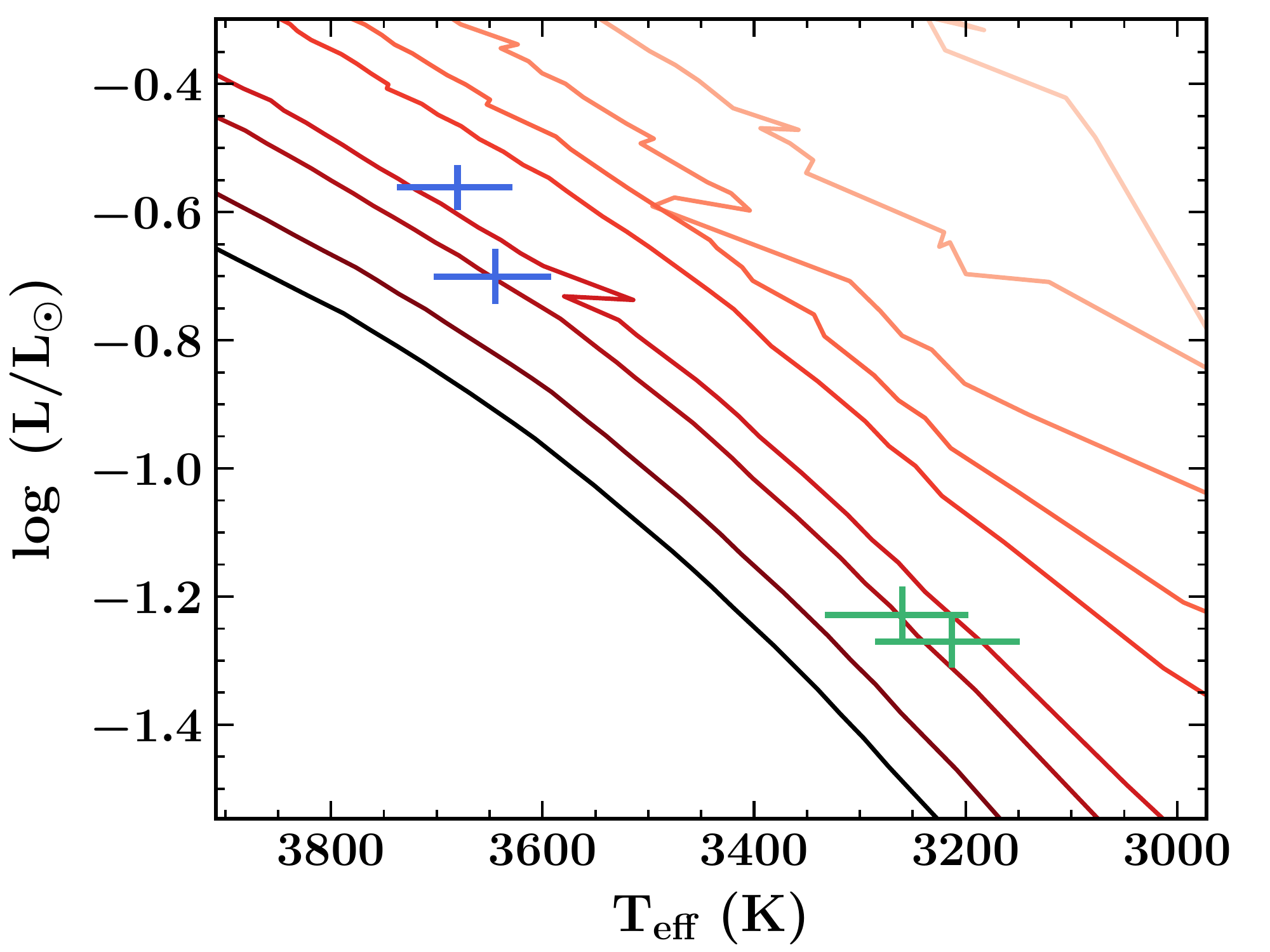} \\
   \includegraphics[width=\psize\linewidth]{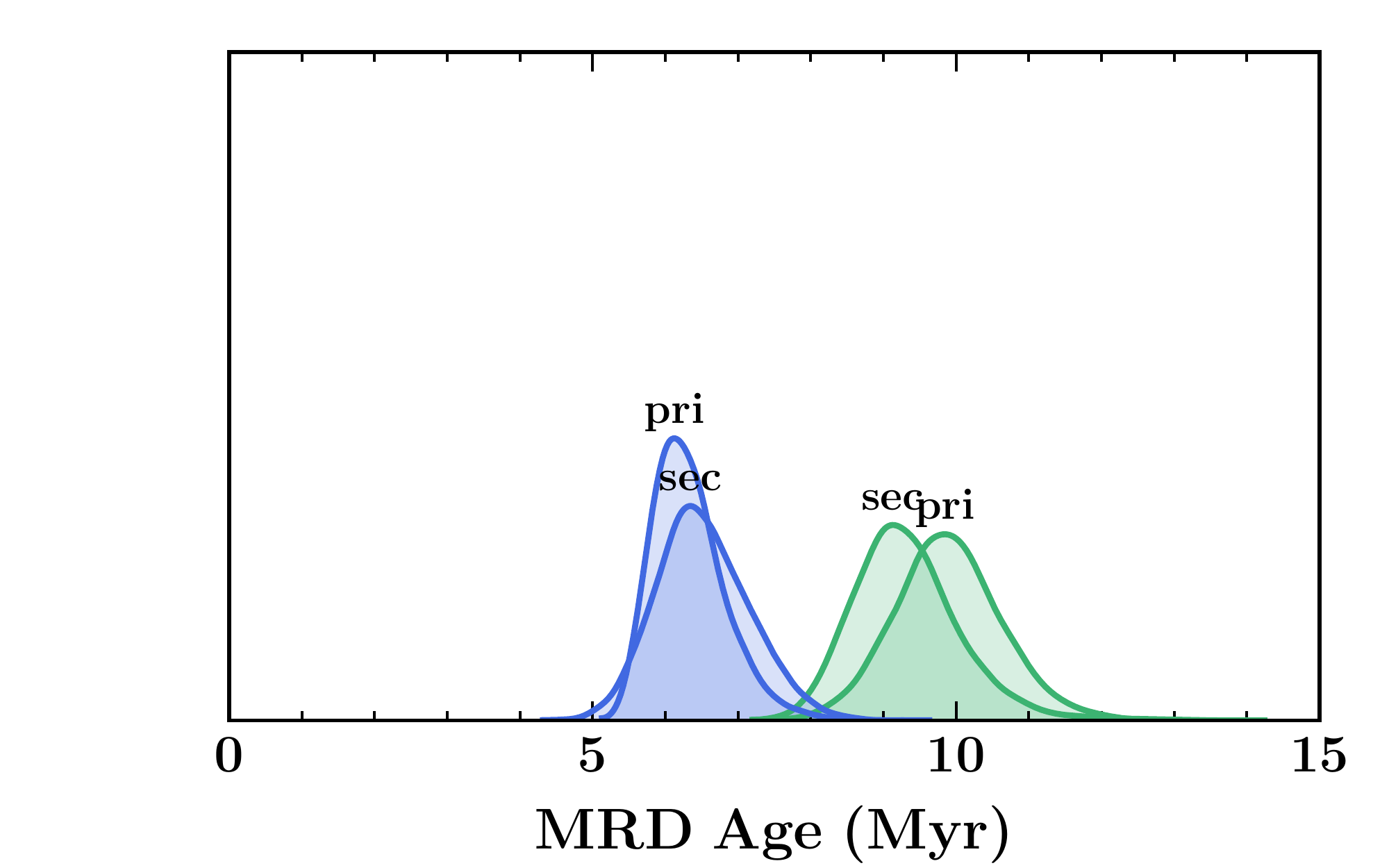}
   \includegraphics[width=\psize\linewidth]{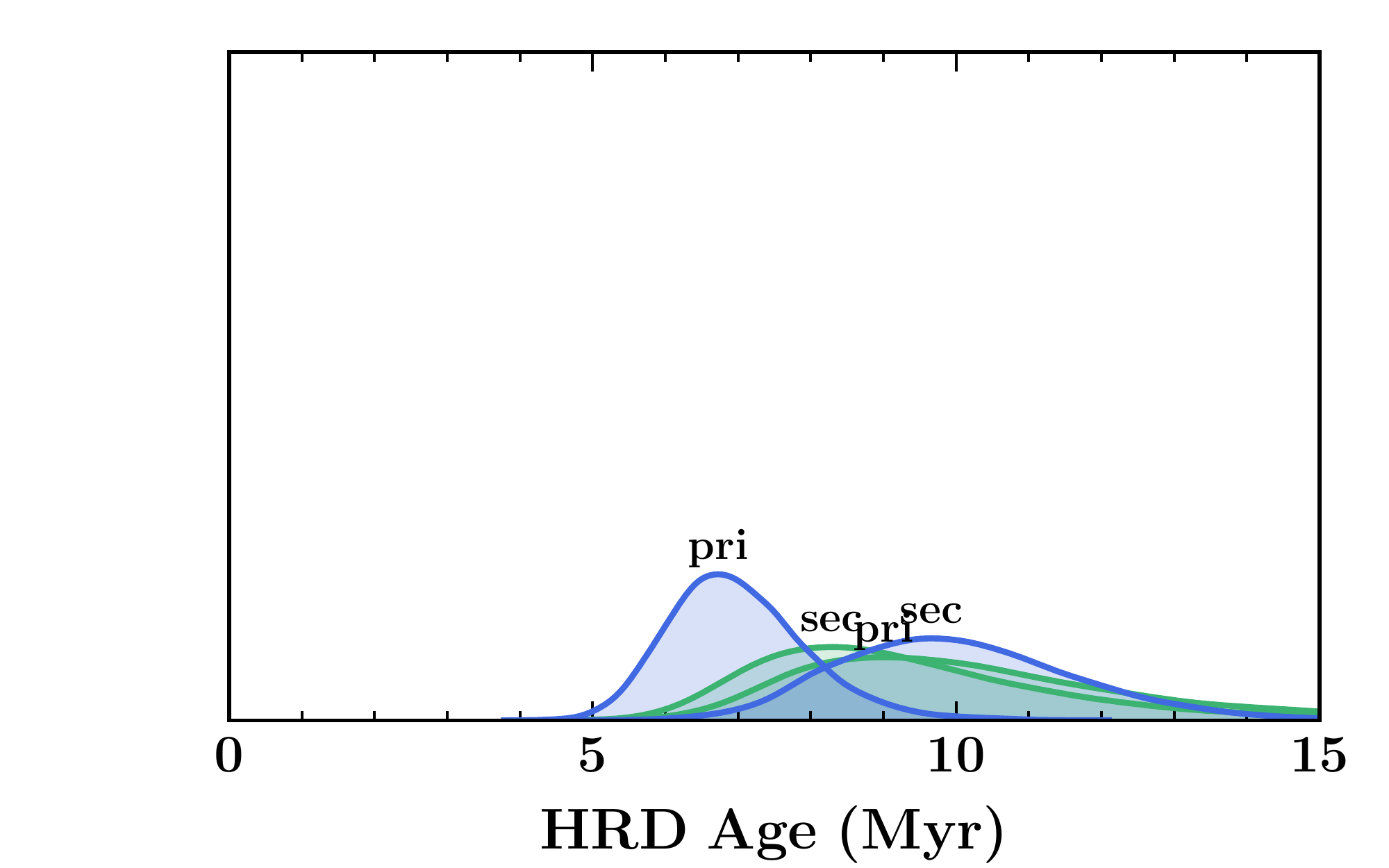} \\
   \caption{Same as Figure \ref{fig:BHAC15_compare} but comparing to the magnetic Feiden models.}
   \label{fig:F16mag_compare}
\end{figure*}

\begin{figure*}
   \centering
   \includegraphics[width=\linewidth]{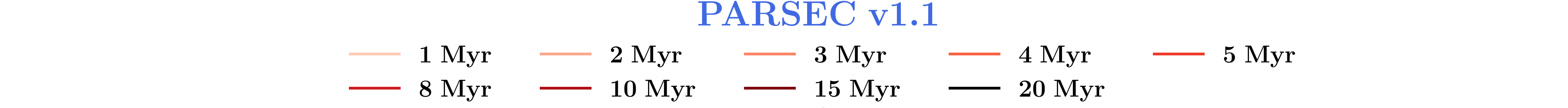}
   \includegraphics[width=\psize\linewidth]{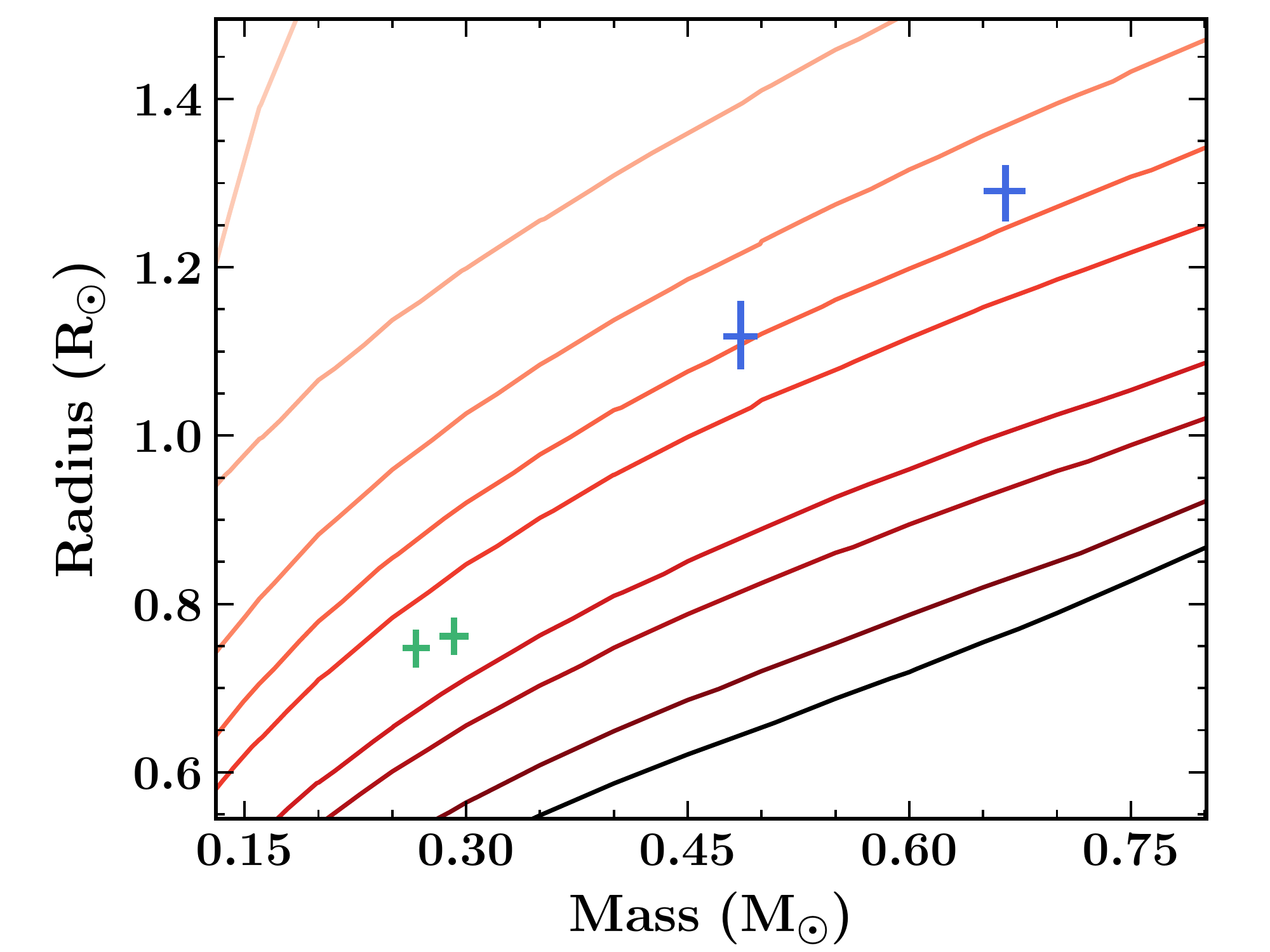}
   \includegraphics[width=\psize\linewidth]{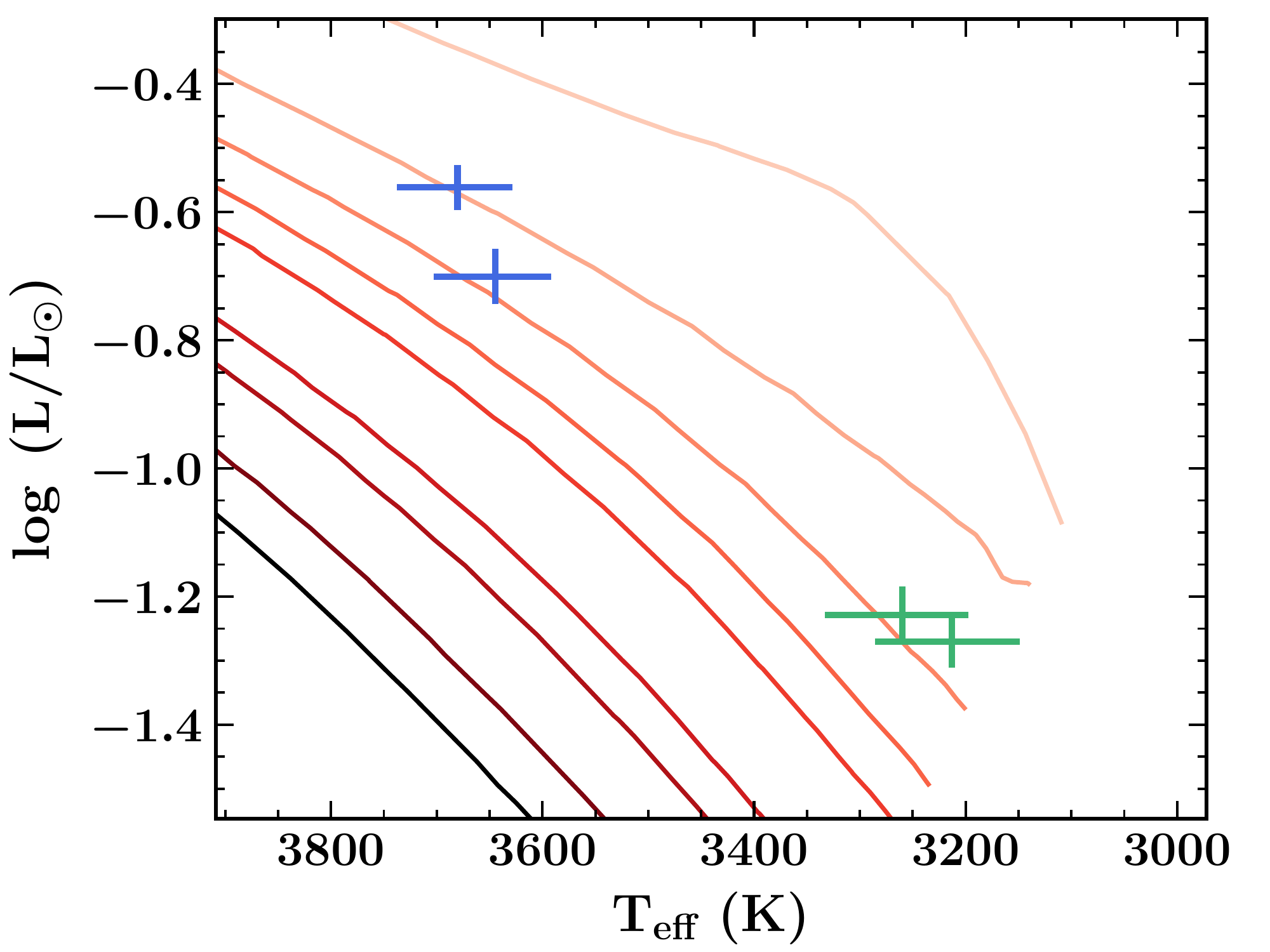} \\
   \includegraphics[width=\psize\linewidth]{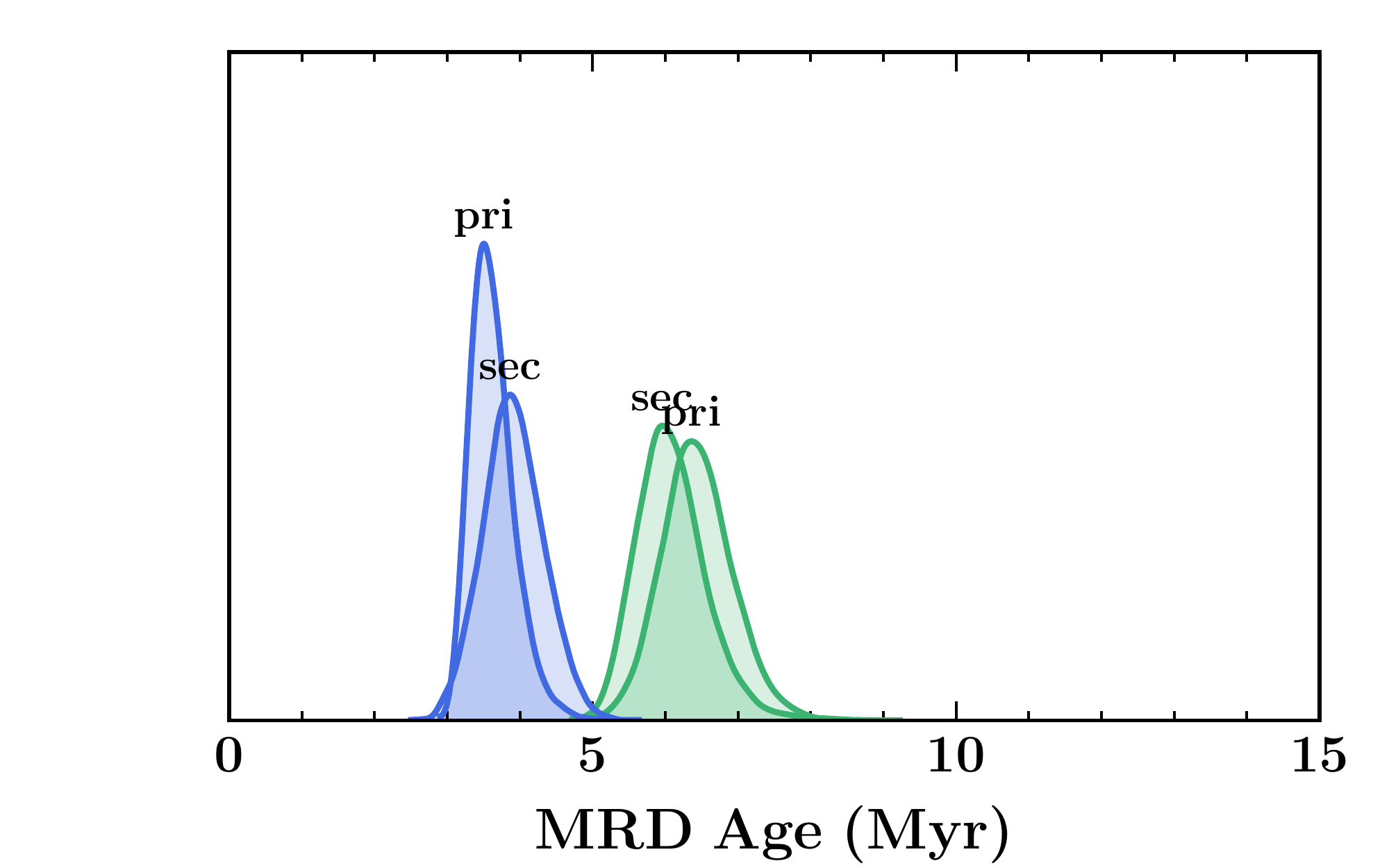}
   \includegraphics[width=\psize\linewidth]{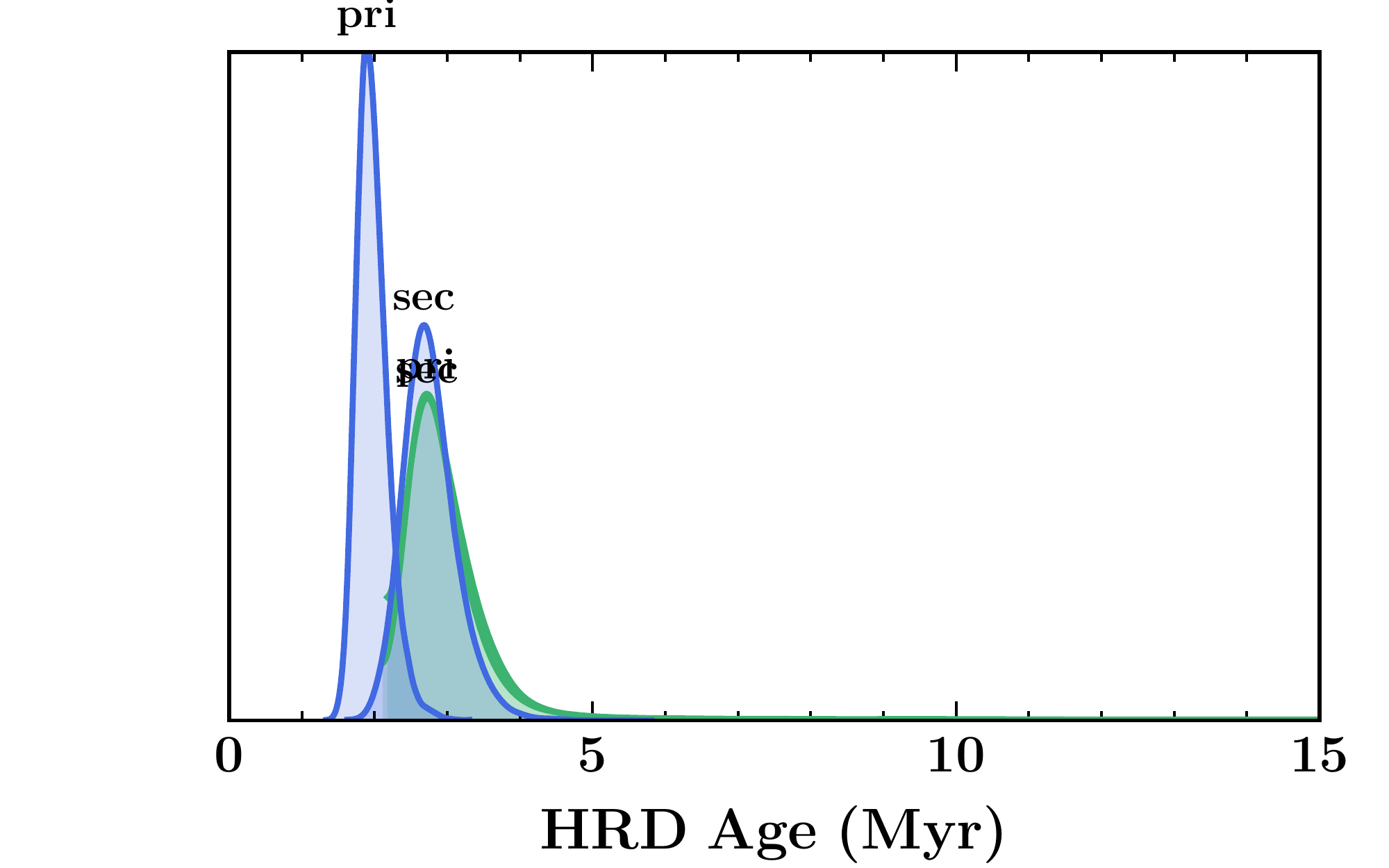} \\
   \caption{Same as Figure \ref{fig:BHAC15_compare} but comparing to PARSEC v1.1 models.}
   \label{fig:Parsec_compare}
\end{figure*}

We simultaneously modelled the \spitzer\ 3.6 and 4.5 \um\ light curves, Keck/HIRES RVs, and the system SED with \gpe. We used 200 `walkers' to step through and explore the parameter space 50,000 times, conservatively discarding the first 30,000 steps as `burn-in', and subsequently thinning each chain following inspection of the autocorrelation lengths for each parameter. 

The model fits to the data are shown in Figures \ref{fig:LCs}--\ref{fig:SED}.
Figure \ref{fig:LCs} shows the \spitzer\ 3.6 and 4.5 \um\ light curves with the global \gpe\ model, which achieves an acceptable fit to the rough out-of-eclipse variability in both bands, as well as the eclipses. Figure \ref{fig:eclipses} shows the detailed fit to the eclipses, which have been phase-folded and detrended with respect to the \gpe\ variability model. The eclipses are grazing and near equal in depth at $\sim$36--37\% for both stars. The secondary-to-primary eclipse depth ratio is slightly higher in the 4.5 \um\ band than in 3.6 \um, which suggests that the secondary star is slightly cooler than the primary. 
The RV orbit solution is displayed in Figure \ref{fig:orbit}, phase-folded on the orbital period. The orbits are close to sinusoidal, suggesting a low eccentricity, and the primary RV semi-amplitude is smaller than the secondary's, indicating a higher primary star mass. The RVs are well-fit by Keplerian orbits with residuals of under a few km\,s$^{-1}$.
The system SED is shown in Figure \ref{fig:SED} which, peaking at $\sim$1--2 \um\ and appearing similar in shape to that of a single star, suggests that both stars are cool and of similar temperature.
The model geometry of the system at primary and secondary eclipse is shown in Figure \ref{fig:geometry}, highlighting that the two stars have similar radii and that the fraction of each star eclipsed is significant. This configuration is required to explain the deep eclipses on both stars.
The main parameters of the fit are given in the top section of Table \ref{tab:params}, with derived parameters in the middle and bottom sections.

We find that \name\ is comprised of two stars with $M_{\rm pri}$ = \Mpri\ \msun\ and $M_{\rm sec}$ = \Msec\ \msun, $R_{\rm pri}$ = \Rpri\ \rsun\ and $R_{\rm sec}$ = \Rsec\ \rsun, and $T_{\rm pri}$ = \Tpri\ K and $T_{\rm sec}$ = \Tsec\ K.
The posterior mass and radius distributions are shown in Figure \ref{fig:MR_compare_SED} (in orange), along with the results from only modelling the light curves and RVs (in grey). Modelling the system SED simultaneously with the light curves and RVs, and hence being able to incorporate the spectroscopic light ratio from the HIRES spectra, has two effects: 
1. the radius uncertainties reduce by a factor of $\sim$3; 
and 2. the radius posteriors change from extended, slightly bimodal, distributions to well-defined unimodal posteriors. 
This improvement primarily results from breaking the degeneracy between the radius ratio and the cosine of the inclination and, to a lesser extent, the surface brightness ratios\footnote{This is smaller because there are two light curve bands, albeit in closely separated bandpasses.}, which is significant when modelling only the light curves and RVs. We feel these improvements justify our approach to model the light curves, RVs, and SED simultaneously. We note that the posterior masses are essentially unchanged, which is reassuring, as we would not expect the inclusion of the SED and spectroscopic light ratio to significantly affect these parameters. Finally, we tested models with constrained GP hyperparameters, such that the \gpe\ variability model was essentially a flat line, and find consistent masses, radii and effective temperatures.


\subsection{Consistent determination of fundamental parameters for CoRoT\,223992193}
\label{sec:EB1039}

In addition to determining fundamental parameters for \name, we also remodelled the other known low-mass EB in \cluster, \ebdisk\ \citep[][hereafter \colblue{G14}]{Gillen14}, with \gpe\ to determine fundamental parameters by simultaneously modelling the \corot\ and \spitzer\ 3.6 \& 4.5 \um\ light curves, VLT/FLAMES and WHT/ISIS RVs, and system SED (see \colblue{G14} and \citet{Gillen17} for details of the data). This was done to obtain a consistent set of parameters that were determined using the same methodology as for \name. These results will be reported in detail in a future publication. Here, we simply report the key parameters: $M_{\rm pri}$ = \Mpriebd\ \msun\ and $M_{\rm sec}$ = \Msecebd\ \msun, $R_{\rm pri}$ = \Rpriebd\ \rsun\ and $R_{\rm sec}$ = \Rsecebd\ \rsun, and $T_{\rm pri}$ = \Tpriebd\ K and $T_{\rm sec}$ = \Tsecebd\ K. We note that these masses and radii are consistent with the values reported in \colblue{G14} to within 1\,$\sigma$ (temperatures were not directly determined in \colblue{G14}). We further note that the masses and radii for both stars are consistent to within 1$\sigma$ both when we do and do not simultaneously model the SED, which gives us confidence that incorporating the SED into the fitting process does not bias our mass and radius measurements.


\section{Discussion}
\label{sec:discussion}

Determining fundamental stellar parameters (masses, radii, and effective temperatures) for stars in EBs provides an observational test of stellar evolution theory. Furthermore, for EBs in young open clusters, the distances determined from their modelling give independent estimates for their host cluster. We present an updated mass-radius relation for detached double-lined EBs in \S \ref{sec:MR_relation} and compare our results for \name\ and \ebdisk\ to the predictions of five stellar evolution models in \S \ref{sec:model_comp}. We then discuss the age and distance of \cluster\ in \S \ref{sec:age} \& \ref{sec:distance}, before comparing the effect of using \btsettl\ or \phoenix\ model atmospheres in \S \ref{sec:BT_vs_PX}.


\subsection{Mass-radius relation for low-mass EBs}
\label{sec:MR_relation}

Figure \ref{fig:MR} shows the mass-radius relation for detached double-lined EBs below 1.5 M$_{\odot}$. The coloured lines represent isochrones from \citet[][hereafter BHAC15]{Baraffe15} from 1\,Myr to 1\,Gyr, and the data points show measurements for EBs in the field (grey) and in sub-Gyr open clusters (coloured, see figure caption for colour scheme). The two EBs in \cluster\ are shown in blue, with the new system presented here, \name, being the lower mass system and \ebdisk\ the higher mass system.


\subsection{Comparison with stellar evolution models}
\label{sec:model_comp}

We compare our mass, radius, and \teff\ measurements (and derived luminosity values) to the predictions of five sets of stellar evolution models in both the \mrp\ and \tlp\ planes. The \mrp\ diagram (hereafter MRD) tests the fundamental (i.e. structural) predictions of the the models while the \tlp\ diagram (i.e. Hertzsprung--Russell, hereafter HRD) tests their predicted radiative properties. We use both \name\ and \ebdisk\ as a joint test of stellar evolution theory because they are both members of \cluster, and hence share the same composition and cluster age.

We compare to five sets of modern stellar evolution models: BHAC15 \citep{Baraffe15}, MIST v1.2 \citep{Dotter16,Choi16}, the standard and magnetic models of \citet[][hereafter Feiden]{Feiden16}, and the PARSEC v1.1 models \citep{Bressan12}.
The BHAC15 models \citep{Baraffe15} are essentially an updated version of the BCAH98 models \citep{Baraffe98}, which most notably now use \btsettl\ model atmospheres and updated surface boundary conditions.
The MIST v1.2 models \citep{Dotter16,Choi16} are based on the MESA (Modules for Experiments in Stellar Astrophysics) stellar evolution package. We use the rotating set of isochrones but note that these are equivalent to the non-rotating versions on the PMS (solid-body rotation is commenced on the Zero Age Main Sequence; ZAMS).
The Feiden models are based on the Dartmouth Stellar Evolution Program \citep[DSEP;][]{Dotter08}, which were further developed in \citet{Feiden12a,Feiden13} and \citet{Feiden16} to include the effect of magnetic fields. Magnetic fields act to inhibit convection and hence slow PMS contraction, which generally results in older age predictions compared to non-magnetic models.
The PARSEC models \citep{Bressan12,Chen14} are available in three versions: v1.0, v1.1 and v1.2S. Both the v1.0 and v1.1 models adopt the grey atmosphere approximation as their external boundary condition, which relates the temperature (T) and the Rosseland mean optical depth ($\tau$) across the atmosphere. The v1.2S models updated the T-$\tau$ relation to use the predictions of the \btsettl\ models, but also included a shift in the T-$\tau$ relation to reproduce the observed mass--radius relation for low-mass dwarf stars. This shift means that the v1.2S models are not a direct test of the underlying stellar evolution theory for our two systems, and hence we compare to v1.1 models. 
We refer the reader to the relevant model papers for further details and to \citet{Stassun14} for a general discussion of stellar evolution models in the context of PMS EBs.

In Figures \ref{fig:BHAC15_compare}--\ref{fig:Parsec_compare}, we compare our results to the BHAC15, MIST v1.2, Feiden (both standard and magnetic), and PARSEC v1.1 stellar evolution models, respectively. The top panels in these figures show \mrp\ and \tlp\ isochrones (coloured lines, left and right plots, respectively), along with the positions of \name\ and \ebdisk\ (in green and blue, respectively). The bottom panels show the model-predicted age distributions for each star based on their positions in the MRD and HRD (left and right plots, respectively). To estimate these ages, we first interpolated the models to compute a finer grid of predictions at each model age, and then compared our posterior distributions for \name\ and \ebdisk\ to this fine grid, which yielded a distribution of model-derived ages. These ages, along with the masses inferred from the positions of the stars in the HRD, are listed in Table \ref{tab:ages}.

From these plots, we can see five noteworthy trends: 
1. \name\ appears older than \ebdisk\ in the MRD for all models; 
2. \name\ appears slightly older than \ebdisk\ in the HRD for the non-magnetic models, but less so than in the MRD (and is comparable for the magnetic Feiden models);
3. the age estimates for both components of each system in the MRD agree well, so while there is an apparent age difference between the two systems, the two stars within each system appear coeval, as we would expect; 
4. the coevality of the components within each system is more complicated in the HRD, where the stars in \name\ appear coeval but for \ebdisk\ the secondary appears older than the primary and in better agreement with \name; 
and 5. the model-derived age distributions are generally tighter from the MRD than the HRD (except for the primary star of \ebdisk, where they are comparable), which results from the measured masses and radii being better constrained than the effective temperatures (and hence also luminosities).

Some subtler trends begin to emerge when we compare the results from different models. The left panel of Figure \ref{fig:frac_compare} shows the fractional difference between the MRD and HRD model ages.
The primary and secondary components of \name\ are shown in blue and orange, and those of \ebdisk\ in pink and gold. The BHAC15, MIST v1.2 and Feiden (std) models (circles, squares and upwards triangles) give very similar predictions for all stars. They find that both components of \name, and the primary of \ebdisk, are on average $\sim$30--35\% ($\sim$2.5 and 1.3 Myr, respectively) younger in the HRD than in the MRD. 
The HRD age of the \ebdisk\ secondary is slightly older than the primary and in agreement with its MRD age (slightly younger on average in the HRD but within or at the 1$\sigma$ level). This results from a higher estimated \teff\ ratio between the two components than the models would expect given their mass ratio.
The PARSEC v1.1 models (diamonds) make predictions that are systematically younger than the other non-magnetic models in both the MRD ($\sim$1.5 and 0.7 Myr for \name\ and \ebdisk), and in the HRD ($\sim$2.4 and 1.3 Myr). These result in a shift leftwards and upwards, respectively, which we see for all four stars in the plot.
The magnetic Feiden models yield ages that are systematically older than the non-magnetic models in both the MRD and HRD. The MRD and HRD ages agree well for both components of \name\ (comfortably within 1$\sigma$), the primary of \ebdisk\ appears $\sim$10\% older in the HRD (but consistent at the 1$\sigma$ level), and the secondary appears $\sim$3 Myr ($\sim$50\%) older in the HRD than the MRD.
Interestingly, all models show the same overall trend: the two components of \name\ and the primary of \ebdisk\ have similar fractional age differences and the secondary of \ebdisk\ sits lower because it has a comparatively older age in the HRD compared to its MRD age. For the non-magnetic models, this brings its MRD and HRD ages into better agreement but into worse agreement for the magnetic Feiden models.

The right panel of Figure \ref{fig:frac_compare} shows the fractional differences between the dynamically-determined masses (from our EB modelling) and the masses inferred from the positions of the stars in the HRD.
This is a similar test to the fractional age differences shown in the left panel, but reframed to investigate the masses we would infer for these stars given their positions in the HRD, and indeed we see similar trends.
For \name\ and the primary of \ebdisk, the masses inferred from the HRD positions of BHAC15, MIST v1.2 and Feiden (std), are on average smaller than the dynamical ones by $\sim$25--35\% ($\sim$0.07 and 0.2 M$_{\odot}$, respectively). 
The dynamical and HRD-inferred masses for the secondary of \ebdisk\ agree to within 1$\sigma$.
The PARSEC v1.1 models give smaller HRD-inferred masses than the other non-magnetic models, which correspond to larger fractional mass differences of $\sim$55\% ($\sim$0.15 M$_{\odot}$) for \name\ and $\sim$30--45\% for \ebdisk\ ($\sim$0.3 and 0.1 M$_{\odot}$ for the primary and secondary, respectively).
The magnetic Feiden models predict larger HRD-inferred masses than the non-magnetic models. These are consistent with our dynamical masses for both components of \name, $\sim$10\% ($\sim$0.05 M$_{\odot}$) larger for the primary of \ebdisk\ (but consistent to within 1$\sigma$), and $\sim$40\% ($\sim$0.2 M$_{\odot}$) larger for \ebdisk's secondary.

For both components of \name\ and the primary of \ebdisk, the magnetic Feiden models perform best with consistent MRD vs. HRD ages and dynamical vs. HRD masses. The properties of \ebdisk's secondary are best matched by the non-magnetic models of BHAC15, MIST v1.2 and Feiden (std). Overall, there is no conclusive trend in the predicted ages and masses for all stars across models.

Finally, we note that both systems do not have known tertiary companions, which is important as \citet{Stassun14} showed that EBs in triple systems are generally less-well explained by current stellar evolution theory than their counterparts without tertiary companions. The dynamical role of a tertiary companion in affecting the evolution of stars in binary configurations is an ongoing question, but one of significant interest \citep[e.g.][]{Gomez-Maqueo-Chew12,Stassun14,Cheng19}.


\subsection{The age of NGC 2264 from EBs}
\label{sec:age}

\cluster\ is typically considered to be $\sim$\,$3^{+5}_{-3}$ Myr \citep{Walker56,Park00,Rebull02}. The two systems presented here appear to lie towards the upper end of this range, with \name\ and \ebdisk\ having apparent ages of $\sim$7--9 and $\sim$4--6 Myr, respectively (from their positions in the MRD). It is interesting to note that \usco\ has typically been considered older than \cluster\ with an age of 5--10 Myr \citep[e.g.][and references therein]{Pecaut12,Pecaut16,Feiden16,Rizzuto16}. Recently, \citet{David19} studied nine EBs in \usco\ and found evidence for a self-consistent age of 5--7 Myr across the 0.3--5 M$_{\odot}$ mass range. Given that the ages of the EBs in \cluster\ and \usco\ are consistent, it is possible that the range of ages spanned by stars in these clusters are more similar than previously recognised. However, it should be kept in mind that the double-lined eclipsing binaries which can be well-measured via both high-precision photometry and precise spectroscopy in order to obtain their radii and masses, may be preferentially sampling the older, disk-free populations in young regions.  Both \cluster\ and \usco\ are adjacent to on-going star formation within molecular cloud cores.


\subsection{The distance to \cluster\ from EBs in the \emph{Gaia} era}
\label{sec:distance}

The distance to \cluster\ has generally been considered to be $\sim$700--800 pc \citep{Sung97,Dahm08,Sung10}, although individual works have suggested a range between $\sim$400 and 900 pc (e.g. \citealt{Dzib14} and \citealt{Baxter09}, respectively). In \colblue{G14} we determined a distance of $d = 756\pm96$ pc to \ebdisk. Subsequently, the \gaia-DR2 parallax measurement yielded a consistent distance estimate of $d = 745^{+83}_{-68}$ pc. From our new modelling, where we simultaneously solve for the distance to the system, we obtain $d =$ \distebd\ pc. For \name\ we determine a distance of $d =$ \dist\ pc, which is in agreement with the \gaia-DR2 estimate of $718^{+67}_{-57}$ pc. Both of our distance estimates are also consistent with \gaia\ DR2-derived distances to \cluster\ of $738^{+23}_{-21}$ pc by \citet{Kuhn19} and $719\pm16$ pc by \citet{Maiz-Apellaniz19}. In our final modelling, we included the \gaia-DR2 parallaxes as priors, but also tested models without them and find consistent distances ($d =$ \distnoplxprior\ pc and $d =$ \distebdnoplxprior\ pc for \name\ and \ebdisk, respectively). This suggests that distances inferred from our modelling are driven by the fit to the light curve, RV and SED data.

\begin{figure*}
   \centering
    \includegraphics[width=\linewidth]{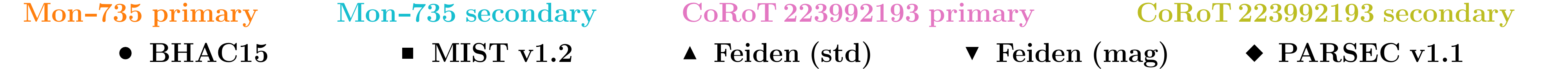}
    \includegraphics[width=0.485\linewidth]{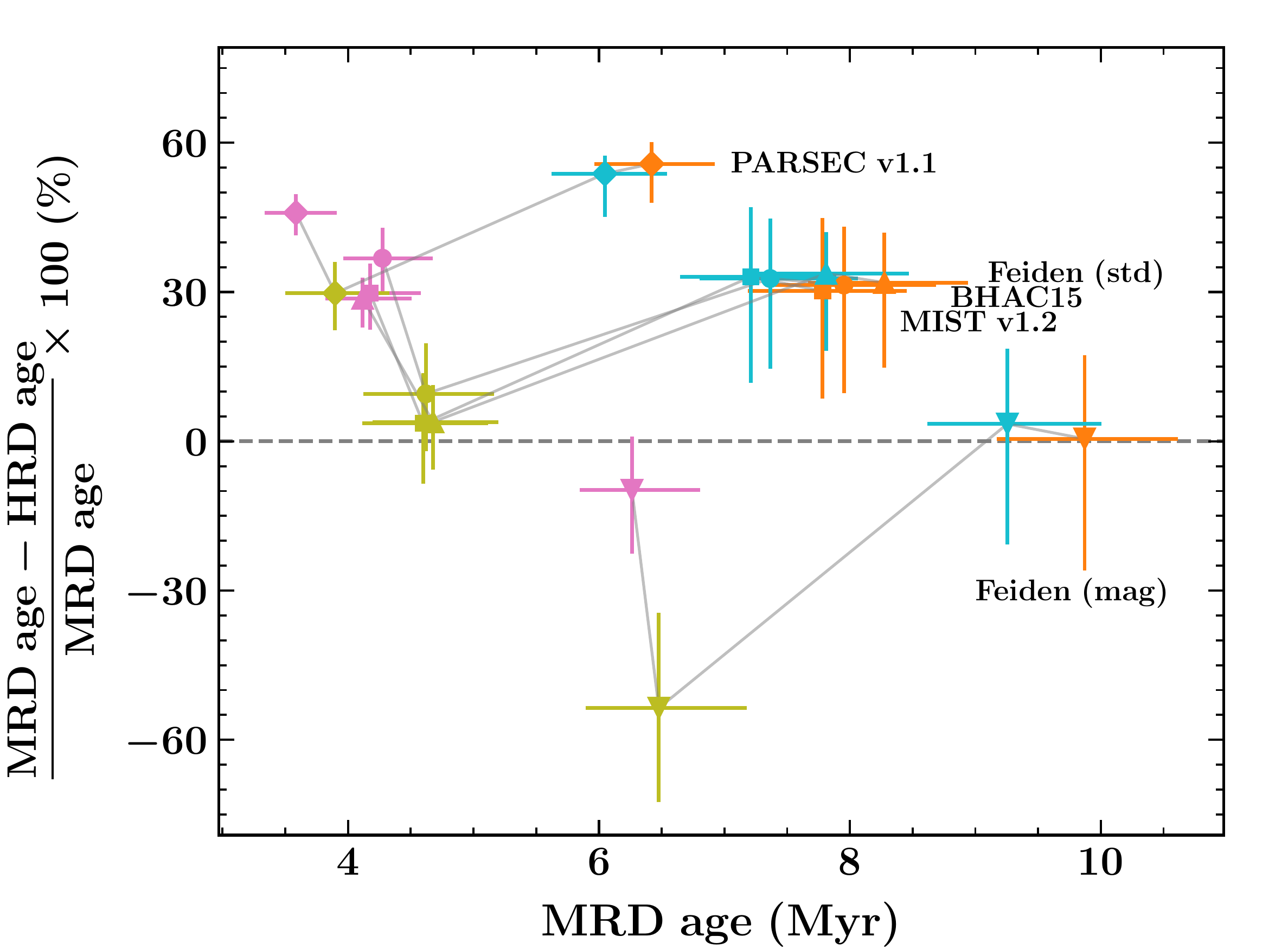}
    \hfill
    \includegraphics[width=0.485\linewidth]{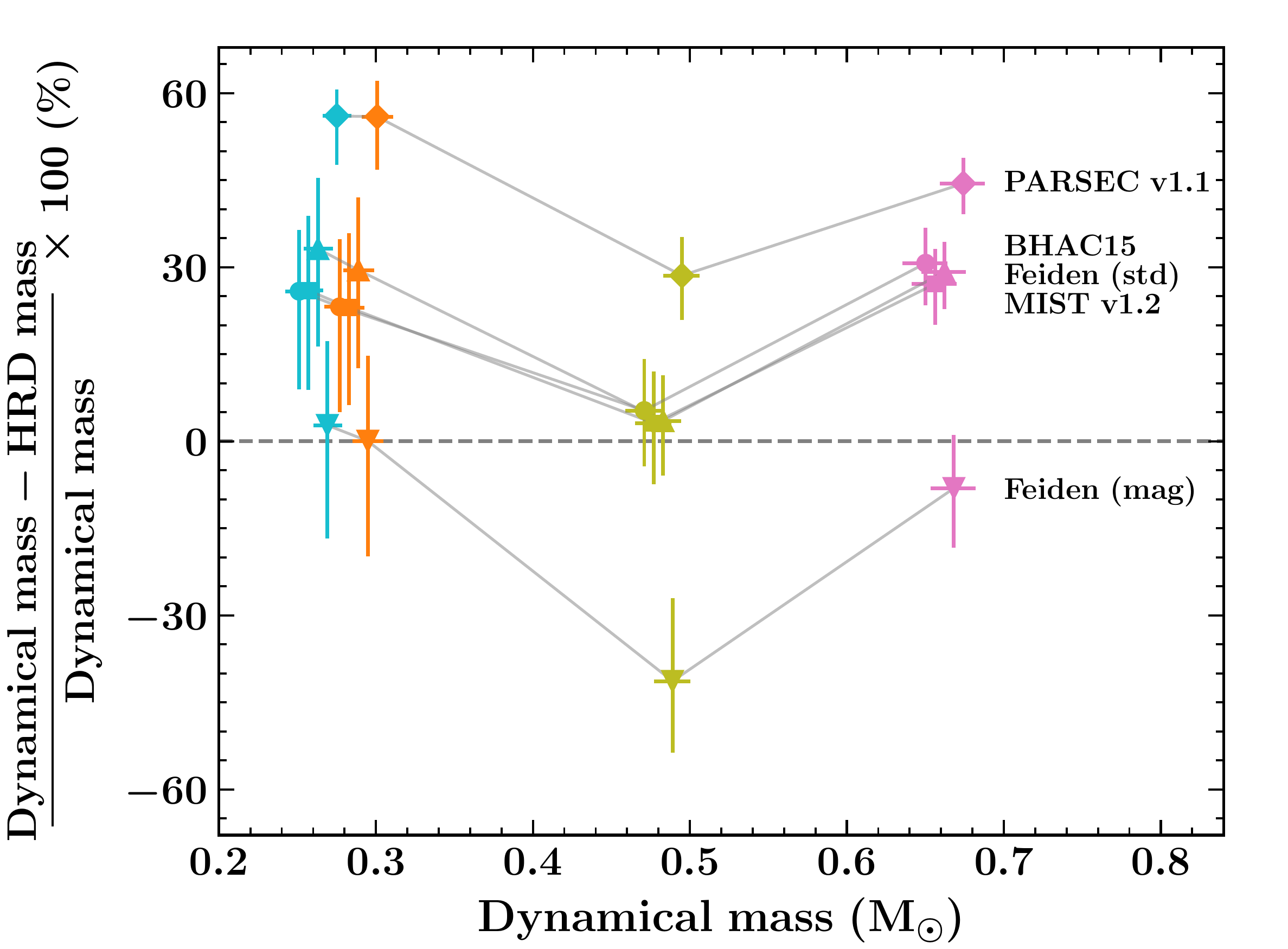}
    \caption{\emph{Left:} Fractional age difference between the MRD and HRD for \name\ and \ebdisk\ according to different PMS stellar evolution models. The colours represent different stars and the symbols different PMS models. \emph{Right:} Fractional mass difference between our dynamically-determined masses and those predicted from the theoretical HRD according to different PMS models. On the x-axis, the different models for each star have been horizontally offset for clarity. The two plots test similar physics and hence show similar (mirrored) trends for the four stars.}
   \label{fig:frac_compare}
\end{figure*}

\begin{table*}
  \centering
  \caption{Isochronal ages of \name\ and \ebdisk\ in the mass-radius diagram (MRD) and Hertzsprung-Russell diagram (HRD), and HRD-inferred masses.}
  \label{tab:ages}
  \begin{tabular}{l@{\hskip 0.5in}rr@{\hskip 0.5in}rr@{\hskip 0.5in}rr}
    \hline
    \hline
    \noalign{\smallskip}
    Model  &  \multicolumn{2}{c}{MRD age (Myr)~~~~~~~~~~~~}  &  \multicolumn{2}{c}{HRD age (Myr)~~~~~~~~~~~~}  &  \multicolumn{2}{c}{HRD mass (\msun)}  \\
      &  Primary  &  Secondary  &  Primary  &  Secondary  &  Primary  &  Secondary \\
    \noalign{\smallskip}
    \hline
    \noalign{\smallskip}
    \multicolumn{7}{c}{\dotfill ~~ \name ~~ \dotfill} \\
    \noalign{\smallskip}

BHAC15        &  $8.0\pm0.7$  &  $7.4\pm0.7$  &  $5.5^{+1.8}_{-1.0}$  &  $5.0^{+1.5}_{-0.9}$  &  $0.22\pm0.05$  &  $0.20\pm0.04$ \\ [\nex]
MIST v1.2     &  $7.8\pm0.7$  &  $7.2\pm0.7$  &  $5.5^{+1.8}_{-1.2}$  &  $4.9^{+1.7}_{-1.1}$  &  $0.22\pm0.04$  &  $0.20\pm0.04$ \\ [\nex]
Feiden (std)  &  $8.3\pm0.7$  &  $7.8\pm0.7$  &  $5.7^{+1.5}_{-0.9}$  &  $5.2^{+1.3}_{-0.7}$  &  $0.21\pm0.05$  &  $0.18\pm0.04$ \\ [\nex]
Feiden (mag)  &  $9.9\pm0.7$  &  $9.3\pm0.8$  &  $9.9^{+2.7}_{-1.8}$  &  $9.0^{+2.5}_{-1.6}$  &  $0.29\pm0.06$  &  $0.26\pm0.05$ \\ [\nex]
PARSEC v1.1   &  $6.4\pm0.5$  &  $6.0\pm0.5$  &  $2.8\pm0.5$          &  $2.8\pm0.5$          &  $0.13\pm0.03$  &  $0.12\pm0.02$ \\

    \noalign{\smallskip} \noalign{\smallskip} 
    \multicolumn{7}{c}{\dotfill ~~ \ebdisk ~~ \dotfill} \\
    \noalign{\smallskip}
    
BHAC15        &  $4.3\pm0.4$  &  $4.6\pm0.5$  &  $2.7\pm0.4$  &  $4.2\pm0.8$  &  $0.46\pm0.05$  &  $0.46\pm0.04$ \\ [\nex]
MIST v1.2     &  $4.2\pm0.4$  &  $4.6\pm0.5$  &  $2.9\pm0.4$  &  $4.4\pm0.8$  &  $0.48\pm0.04$  &  $0.47\pm0.05$ \\ [\nex]
Feiden (std)  &  $4.1\pm0.4$  &  $4.7\pm0.5$  &  $3.0\pm0.4$  &  $4.5\pm0.7$  &  $0.47\pm0.04$  &  $0.47\pm0.04$ \\ [\nex]
Feiden (mag)  &  $6.3\pm0.5$  &  $6.5\pm0.7$  &  $6.9\pm1.0$  &  $9.9\pm1.7$  &  $0.72\pm0.06$  &  $0.69\pm0.07$ \\ [\nex]
PARSEC v1.1   &  $3.6\pm0.3$  &  $3.9\pm0.4$  &  $1.9\pm0.2$  &  $2.7\pm0.4$  &  $0.37\pm0.03$  &  $0.35\pm0.04$ \\
  
    \noalign{\smallskip}
    \hline
 \end{tabular}
\end{table*}


\subsection{Differences between using \btsettl\ and PHOENIX atmosphere models}
\label{sec:BT_vs_PX}

For the main results presented here, we opted to use \btsettl\ model atmospheres in our SED modelling instead of \phoenix\ v2 models. This was primarily because the \btsettl\ models yielded distances that were in better agreement with \gaia, but also because they achieved a slightly better fit to the SED of \name\ (although the \phoenix\ fit was still acceptable). 

To test the effect of the stellar atmosphere model adopted, we  performed additional modelling runs for both \name\ and \ebdisk\ with the \phoenix\ models. Using the \phoenix\ models for both \name\ and \ebdisk\ resulted in essentially unchanged masses and radii (easily consistent to within 1$\sigma$). The \teff's, however, were lower by $\sim$\teffBTPHdiff\,K for \name\ and $\sim$\ebdteffBTPHdiff\,K for \ebdisk. Correspondingly, the distances were smaller by $\sim$\distdiff\ and $\sim$\ebddistdiff\ pc, respectively, which are both smaller than the \gaia\ distances but consistent to within 2$\sigma$.

With lower \teff's, the \phoenix\ models suggest the stars are younger in the HRD by $\sim$1--3 Myr for \name\ and $\sim$0.5--2 Myr for \ebdisk\ (compared to the \btsettl\ models). The positions in the MRD are essentially unchanged. In the fractional age and mass plots (Figure \ref{fig:frac_compare}), therefore, the values for each star shift upwards by $\sim$20--30\% for \name\ and $\sim$10--30\% for \ebdisk, while the overall trend between stars remains similar\footnote{The PARSEC v1.1 models do not extend down to the PHOENIX-derived \teff's for \name\ and hence were not included in this comparison.}.

\section{Conclusions}
\label{sec:conclusions}

We have presented the discovery and characterisation of \name\ as a detached, double-lined PMS EB member of the \cluster\ star forming region. \name\ possesses IR \spitzer\ discovery light curves, as well as optical follow-up Keck/HIRES RVs. As the system displays relatively equal depth eclipses, modelling only the IR light curves and optical RVs suffered from a degeneracy between the radius ratio and inclination, and to a lesser extent the surface bright ratios, which led to large radius uncertainties. We overcame this issue by simultaneously modelling the system SED along with the light curves and RVs, which allowed us to apply a light ratio constraint from the HIRES spectra and propagate it through the SED model into the \spitzer\ light curve bands and hence into the eclipse modelling; this broke the aforementioned degeneracy and yielded well-constrained radii. As a result of this effort, \gpe\ now has the capability to simultaneously model light curves, RVs and SEDs.

Simultaneously modelling the \spitzer\ light curves, Keck/HIRES RVs, and system SED, we determine self-consistent masses, radii and temperatures for both component stars, as well as the distance to \name\ and, by extension, \cluster. We find that the two stars have masses of $M$ = \Mpri\ and \Msec\ M$_{\odot}$, radii of $R$ = \Rpri\ and \Rsec\ R$_{\odot}$, and effective temperatures of \teff\ = \Tpri\ and \Tsec\ K, and travel on circular orbits around their common centre of mass in $P$ = \period\ days. In addition to \name, we also remodelled \ebdisk, the other known low-mass EB in \cluster, to derive a consistent set of parameters for this system, so we could make a joint comparison for both systems to the predictions of different stellar evolution models. 

We compare to the BHAC15, MIST v1.2, Feiden (both standard and magnetic), and PARSEC v1.1 models and find that the lower mass \name\ appears older than \ebdisk\ in the MRD and, to a lesser extent, in the HRD. In the MRD, \name\ appears to be $\sim$7--9 Myr while \ebdisk\ appears to be $\sim$4--6 Myr. The two components in each EB system possess consistent MRD ages, whereas in the HRD, the components of \name\ are coeval but the secondary of \ebdisk\ appears slightly older than the primary, and in better agreement with \name. For both components of \name\ and the primary of \ebdisk, the magnetic Feiden models perform best with consistent MRD vs. HRD ages and dynamical vs. HRD masses. The MRD and HRD properties of \ebdisk's secondary are best matched by the non-magnetic models of BHAC15, MIST v1.2 and Feiden (std). For these two low-mass PMS systems in \cluster, there is no clear trend between models in either the MRD- and HRD-derived ages or the HRD-inferred masses, except that the magnetic models predict older ages and higher masses than the non-magnetic models.

\ebdisk\ and \name\ are the first two low-mass EBs to come out of the CoRoT and \spitzer\ observations of \cluster, with more systems in preparation. These will form a powerful sample of near-coeval EB systems, formed from the same parent molecular cloud, with which to test PMS stellar evolution theory and better understand both the age of, and age spread within, the \cluster\ region.

\section*{Acknowledgements}

The authors thank Hannu Parviainen, Trevor David, Grant Kennedy and Didier Queloz for helpful discussions, and the anonymous referee for their insightful reading of the manuscript and helpful suggestions for improvement.
EG gratefully acknowledges support from the David and Claudia Harding Foundation in the form of a Winton Exoplanet Fellowship.

This work is based in part on observations made with the \spitzer\ Space Telescope, which is operated by the Jet Propulsion Laboratory, California Institute of Technology under a contract with NASA.
Some of the data presented herein were obtained at the W. M.
Keck Observatory, which is operated as a scientific partnership
among the California Institute of Technology, the University of
California, and the National Aeronautics and Space Administration. 
The Observatory was made possible by the generous
financial support of the W. M. Keck Foundation. The authors
wish to recognize and acknowledge the very significant cultural
role and reverence that the summit of Maunakea has always
had within the indigenous Hawaiian community. We are most
fortunate to have the opportunity to conduct observations from
this mountain.
This research has made use of the VizieR catalogue access tool, CDS, Strasbourg, France (DOI : 10.26093/cds/vizier). The original description of the VizieR service was published in A\&AS 143, 23 \citep{Ochsenbein00}.
This research has made use of the SIMBAD database, operated at CDS, Strasbourg, France \citep{Wenger00}.
This research has made use of the SVO Filter Profile Service (FPS; http://svo2.cab.inta-csic.es/theory/fps/) supported from the Spanish MINECO through grant AYA2017-84089, and described in \citet{Rodrigo12}.
This publication makes use of data products from the Two Micron All Sky Survey, which is a joint project of the University of Massachusetts and the Infrared Processing and Analysis Center/California Institute of Technology, funded by the National Aeronautics and Space Administration and the National Science Foundation.
This publication makes use of data products from the Wide-field Infrared Survey Explorer, which is a joint project of the University of California, Los Angeles, and the Jet Propulsion Laboratory/California Institute of Technology, funded by the National Aeronautics and Space Administration.




\bibliographystyle{mnras}
\bibliography{ref}






\bsp	
\label{lastpage}
\end{document}